\documentclass{article}
\usepackage{mathtools, amssymb, mathrsfs, amsthm}
\usepackage[margin=1in]{geometry}
\usepackage{graphicx}
\graphicspath{{images/}}
\usepackage{caption}
\usepackage{subcaption}
\usepackage{enumitem}
\usepackage{titlesec}
\usepackage[utf8]{inputenc}
\usepackage[english]{babel}
\usepackage{ying}
\usepackage[font = small]{caption}
\usepackage[colorlinks = true, citecolor = blue, urlcolor = blue]{hyperref}
\usepackage{xcolor}
\usepackage{cleveref}
\usepackage{caption}
\usepackage{subcaption}
\usepackage{sectsty}

\usepackage{parskip}
\newtheorem{lemma}{Lemma} 
\newtheorem{proposition}{Proposition}

\newtheorem{assumption}{Assumption}
\theoremstyle{definition}  
\newtheorem{remark}{Remark}
\newtheorem{example}{Example}

\newcommand{\R}{\mathbb{R}}
\newcommand{\D}{\mathcal{D}}
\renewcommand{\P}{\textup{Pr}}

\newcommand{\Eone}{\E_{\D_1}}
\newcommand{\Etwo}{\E_{\D_2}}
\DeclareMathOperator*{\argmin}{argmin}

\usepackage{tikz}
\usetikzlibrary{positioning}

\begin{document}

\title{Diagnosing the role of observable distribution shift\\ in scientific replications}
\author{Ying Jin\thanks{Equal contribution.}  \and 
Kevin Guo$^*$  \and 
Dominik Rothenh\"{a}usler}
\date{Department of Statistics, Stanford University \\[2ex]
\today}
\maketitle

\begin{abstract}
Many researchers have identified distribution shift as a likely contributor to the reproducibility crisis in behavioral and biomedical sciences.  The idea is that if treatment effects vary across individual characteristics and experimental contexts, then studies conducted in different populations will estimate different average effects.  This paper uses ``generalizability" methods to quantify how much of the effect size discrepancy between an original study and its replication can be explained by distribution shift on observed unit-level characteristics.  More specifically, we decompose this discrepancy into ``components" attributable to sampling variability (including publication bias), observable distribution shifts, and residual factors.  We compute this decomposition for several directly-replicated behavioral science experiments and find little evidence that observable distribution shifts contribute appreciably to non-replicability.  In some cases, this is because there is too much statistical noise.  In other cases, there is strong evidence that controlling for additional moderators is necessary for reliable replication.
\end{abstract}

\section{Introduction}

\subsection{Overview} \label{section:overview}

Several scientific disciplines have recently conducted large-scale reproducibility projects wherein independent laboratories attempt to replicate\footnote{In this paper, we use the term ``replication" to refer to a follow-up experiment with new data, rather than a re-analysis of the original data.} key findings in the field.  Prominent examples include projects in psychology \citep{reproducibility_project_psychology}, experimental economics \citep{camerer2016evaluating}, experimental philosophy \citep{cova2021estimating}, operations management \citep{davis_etal2022} and pre-clinical cancer biology \citep{errington_etal2021}.  

A common finding in these projects is that many results fail to replicate in follow-up experiments.  In Reproducibility Project: Psychology, only 39 out of 100 replications were deemed successful by their administrators \citep{reproducibility_project_psychology}.  In Reproducibility Project: Cancer Biology, only 51 out of 112 studies satisfied three or more objective criteria for replicability \citep{errington_etal2021}.  Even among successful replications, estimated effect sizes tend to vary substantially across original and replication experiments. 

Many statisticians and scientists have identified distribution shift as a likely contributor to these low replication rates, alongside publication bias and low statistical power.  The idea is that treatment effects which vary across individual characteristics and experimental contexts might be harder to detect in some populations than others.  For example, \cite{gelman2015connection} writes ``If effects are different in different places and at different times, then episodes of nonreplication are inevitable, even for very well-founded results."  \cite{bryan2021behavioural} give a concrete example where an intervention designed to reduce energy consumption showed promise on high-income households, but effect sizes declined in broad-based follow-up experiments.  They explain ``Average effects [will] frequently ... be smaller in later-conducted studies even in the absence of type-I error ... because researchers tend to conduct initial studies in samples and contexts that are optimized for effects to emerge."

The main idea of this paper is to use statistical methods from the ``generalizability" literature to estimate \emph{how much} of the effect-size discrepancy in an experiment-replication pair is attributable to distribution shift on observed unit-level characteristics.  Such estimates may improve our scientific understanding, guide reporting, or inform future data-collection efforts \citep{simons2017constraints}.  For instance, if a large part of the discrepancy is explainable by observed distribution shifts, researchers may wish to report heterogeneous treatment effects that vary with these characteristics.  On the other hand, if little-to-no observed discrepancy is attributable to observed distribution shifts, then we may need theory to suggest additional moderating variables that inform how far the original effect generalizes.  We emphasize that our goals are distinct from works seeking to \emph{detect} treatment effect heterogeneity or distribution shift \citep{ding2019decomposing, smith2023prediction}.  Even if these are present, they may not contribute to the discrepancy of interest.

Our specific proposal uses unit-level data from an original study and its direct replication to decompose the discrepancy in their effect estimates into several interpretable components: 

\begin{itemize}
\item \emph{Covariate shift}.  Covariate shift is the part of the observed discrepancy that can be attributed to differences in measured background characteristics.  Such differences are widely cited by both replicators and original study authors.  For example, \cite{flynn2020failure} conjecture that their non-replication of \cite{wood2009positive} could be due to their sample's ``higher number of women than the original study." Similarly, \cite{buttrick_etal} cite the ``significant increase in median age ... compared to the first round of data collection" in their replication report.

\item \emph{Mediation shift}.  Mediation shift captures the discrepancy attributable to the heterogeneous response of key intermediary variables, that is, variables that can be impacted by the treatment. For example, if a biological treatment affects outcomes by increasing production of a certain protein, then a replication experiment where the treatment fails to induce the same level of protein production (e.g. because of unfavorable ambient conditions) may estimate a smaller effect, even if the protein-outcome relationship is actually constant across cell types and environments. 

\item \emph{Sampling variability}.  Even in the absence of distribution shift, experimental estimates may differ simply because of chance factors in sample selection or treatment assignment.  In addition, original effect sizes may be inflated due to selection bias by journals or replicators.  This introduces systematic sampling error that goes above and beyond purely i.i.d.~sampling variability.

\item \emph{Residual factors}.  Finally, differences may stem from subtle differences in treatment implementation, the distribution of hidden moderators, or (apparently) spurious conditions like weather or experimenter identity \citep{linden2021heterogeneity, bryan2021behavioural}. 
 
\end{itemize}

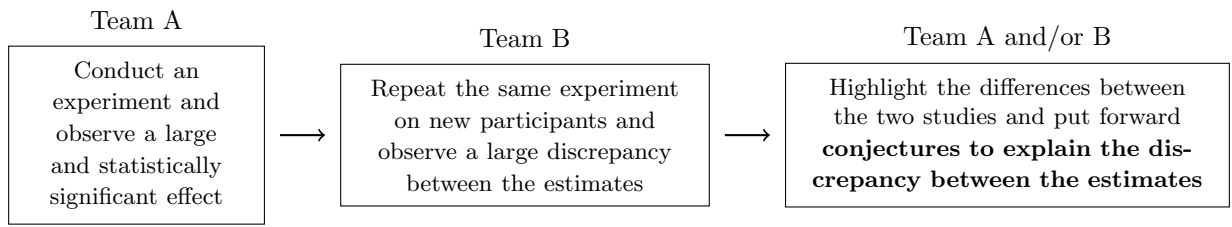
\begin{figure}
\centering
\begin{tikzpicture}
    \node[draw, rectangle, minimum width=2cm, text width=3cm, minimum height=1cm, align = center, inner sep=2mm] (box1) {{\small Conduct an experiment and observe a large and statistically significant effect 
    }};
    \node[draw, rectangle, minimum width=2cm, text width=4.5cm, minimum height=1cm, align = center, inner sep=2mm, right=of box1] (box2) {{\small Repeat the same experiment on 
 new participants and observe a large discrepancy between the estimates 
 }
 };
    \node[draw, rectangle, minimum width=2cm, text width=5.5cm, minimum height=1cm, align = center, inner sep=2mm, right=of box2] (box3) {{\small Highlight the differences between the two studies and put forward \\ \textbf{conjectures to explain the discrepancy between the estimates}}};
    
    \node[above=of box1, yshift=-0.9cm] {Team A};
    \node[above=of box2, yshift=-0.9cm] {Team B};
    \node[above=of box3, yshift=-0.98cm] {Team A and/or B};
    
    \draw[->, thick, shorten >=2mm, shorten <=2mm] (box1) -- (box2);
    \draw[->, thick, shorten >=2mm, shorten <=2mm] (box2) -- (box3);
    
\end{tikzpicture}
\caption{Typical flow of the scientific discussion between the original research team and the replication team. The goal of the discrepancy decomposition introduced in this paper is to rigorously evaluate the testable conjectures advanced in the third panel.
} \label{fig:replication_flowchart}
\end{figure}

See \Cref{fig:replication_flowchart} for an example of how this decomposition could inform scientific discourse.  Our approach is complementary to analyses based on meta-regression~\citep{borenstein2021introduction}.  While these analyses give more insight into the ``Residual factors" component and do not require unit-level data, they are not implementable until a large number of studies on nearly the same phenomenon have been conducted.  In contrast, our methods can be applied when only two studies' data are available.

We compute such a discrepancy decomposition for several directly-replicated experiments in behavioral science.  Perhaps surprisingly, our analysis reveals that the contribution of distribution shift on observed covariates and mediators is often quite small.  Thus, even if substantial heterogeneity is present, the variables measured as part of routine experimental practice do not seem to capture much of it.  This finding is broadly consistent with the findings in the first three Many Labs projects in social psychology \citep{manylabs1, manylabs2, manylabs3} and other meta-analyses \citep{linden2021heterogeneity, rott2023causally}.  For example, the authors of the Many Labs 2 project wrote: ``for the variation investigated here, heterogeneity across samples does not provide much explanatory power for failures to replicate" \citep{manylabs2}.

From a methodological perspective, there has been much progress in terms of dealing with covariate shift and selection bias. Our results suggest that additional methodological progress is needed to further account for potential changes in residual factors.

\subsection{Data and code}

An \textsf{R} package and Shiny application implementing the methods in this article can be found at \url{https://github.com/ying531/repDiagnosis}.  As a companion to this package, we have also compiled a repository of publicly available datasets for directly-replicated experiments at \url{https://github.com/ying531/awesome-replicability-data}.  It contains 11 sets of studies with unit-level covariate information, which we hope will be useful to other researchers investigating the role of distribution shift in scientific replicability.

\subsection{Setting}

The methods in this paper apply to pairs of randomized experiments with published unit-level data.  We denote the available datasets by $\mathcal{D}_1$ and $\mathcal{D}_2$:
\begin{align*}
\mathcal{D}_1 &= \{ (X_i, T_i, M_i, Y_i) \}_{1 \leq i \leq n_1} &\text{``Original" dataset}\\
\mathcal{D}_2 &= \{ (X_j, T_j, M_j, Y_j) \}_{1 \leq j \leq n_2} &\text{``Replication" dataset}.
\end{align*}
In both datasets, $X \in \mathcal{X}$ represents a vector of background characteristics, $T \in \{ 0, 1 \}$ a randomized treatment variable, $M \in \mathcal{M}$ a vector of candidate mediators\footnote{Although we use causal language in this paper for interpretability, we do not make formal causal assumptions.  For example, our methods will remain valid even if $M$ is not actually a mediator in the causal sense.}, and $Y \in \R^p$ a collection of $p \geq 1$ outcome measurements.  Although $p = 1$ is typical, allowing $p > 1$ is needed to analyze experiments with repeated measurements or unit-level clustering.  We analyze such experiments in Sections \ref{section:emotion_time_preference} and \ref{section:covid_misinfo}.  Either $X$ or $M$ is allowed to be empty, to capture problems where no covariates or no mediators are measured.

We require that both studies use the same statistical method to estimate the (one-dimensional) ``effect" of interest.  More specifically, we assume that the original estimate $\theta(\D_1)$ and the replication estimate $\theta(\D_2)$ are obtained from the following ordinary least-squares (OLS) regression model:
\begin{align*}
(\theta(\D_k), \beta(\D_k)) &= \argmin_{(\theta, \beta) \in \R^{\textup{dim}(f) + \textup{dim}(g)}} \sum_{i \in \D_k} \sum_{\ell = 1}^p [ Y_{i \ell} - (\theta, \beta)^{\top} ( T f_{\ell}(X_i), g_{\ell}(X_i)) ]^2.
\end{align*}
Here, $f_{\ell} : \mathcal{X} \to \R^{\textup{dim}(f)}$ and $g_{\ell} : \mathcal{X} \to \R^{\textup{dim}(g)}$ are some known functions, with $f_{\ell}$ containing an intercept and $g_{\ell}$ containing $f_{\ell}$ as a sub-vector.  This covers many commonly-used analyses, such as:
\begin{itemize}[itemsep=-0.5ex]
    \item \emph{Two-sample $t$-test}.  When $p = 1$, setting $f_1(x) \equiv g_1(x) \equiv 1$ makes $\theta(\D_k)$ the difference-of-means estimate used in the (unpaired) two-sample $t$-test.
    \item \emph{Two-way ANOVA}.  If the outcomes $\{ Y_{i \ell} \}_{\ell \leq m}$ are measurements under different settings of a secondary factor $\ell$, then setting $f_{\ell}(x) \equiv 1, g_{\ell}(x) = (1, \mathbf{1} \{ \ell = 1 \}, \dots, \mathbf{1} \{ \ell = m \})$ makes $\theta(\D_k)$ the main effect of $T$ in a two-way analysis of variance model without interactions. 
    \item \emph{Analysis of covariance}.  For a single outcome measurement, setting $f_1(x) \equiv 1$ and $g_1(x) = (1, x)$ makes $\theta(\D_k)$ the main effect of $T$ in a regression of $Y$ on $T$ and $X$.
    \item \emph{Linear covariate adjustment in causal inference}. For $p=1$, when $X$ is mean-centered in each dataset, taking $f_1(x)=(1,x)$ and $g_1(x)=(1,x)$ makes $\theta(\cD_k)$ the covariate-adjusted estimator for the average treatment effect. 
\end{itemize}

We assume that the observations in $\D_1$ and $\D_2$ are samples from populations $P$ and $Q$, respectively.  This assumption is generally satisfied for experiments conducted on online experimentation platforms, but may be less credible for laboratory experiments conducted on volunteer samples.  

We impose the following conditions on these populations.

\begin{assumption} \label{assumpton:primitives}
The following conditions hold:
\begin{enumerate}[itemsep=-0.5ex, topsep=-0.5ex, label=(A\arabic*)]
\item \emph{Design}. Both studies use the same Bernoulli experimental design, where the treatment $T$ is randomized independently of background characteristics $X$. \label{item:rct}
\item \emph{Covariate overlap}. For some $\epsilon > 0$ and all events $A \subset \mathcal{X}$, $Q(X \in A) \geq \epsilon P(X \in A)$.\label{item:covariate_overlap}
\item \emph{Mediator overlap}. For some $\epsilon > 0$ and all events $A \subset \mathcal{M}$, 
$Q(M \in A \mid T, X) \geq \epsilon P(M \in A \mid T, X) $. \label{item:mediator_overlap}
\end{enumerate}
\end{assumption}

Covariate overlap requires that any background characteristics present in the original population can also be observed in the replication population.  If this condition is violated (e.g. if there are women in the original experiment but no women in the replication experiment), then no statistical method can estimate the differences attributable to covariate shift.  This assumption constrains what variables can be included in $X$.  For example, ``country" cannot be part of $X$ if the original and replication experiments were conducted in different countries.  Mediator overlap requires that any treatment-mediator response observed in the original population can be observed in the replication population.  In the case where no covariates (resp. mediators) are measured, Assumption \ref{item:covariate_overlap} (resp. Assumption \ref{item:mediator_overlap}) is satisfied vacuously.
 
\subsection{Notation}

This paper uses the following notation.  For sub-vectors $V, U$ of $(X, T, M, Y)$, $P_V$ and $Q_V$ denote the marginal distribution of $V$ under $P$ and $Q$, respectively.  Meanwhile, $P_{U \mid V}$ and $Q_{U \mid V}$ denote the conditional distributions.  The notation $P_U \otimes Q_{V \mid U}$ denotes the distribution which first samples $U \sim P_U$ and then, given $U = u$, samples $V \sim Q_{V \mid U}(\cdot \mid U = u)$.  Formally, for any subsets $\cU$ and $\cV$ of the sample spaces, $(P_U \otimes Q_{V \mid U})\{ U \in \mathcal{U}, V \in \mathcal{V} \} = \E_{P_U}[ Q_{V \mid U}( \mathcal{V} \mid U) \mathbf{1} \{ U \in \mathcal{U} \}]$.

For any distribution $F$ on $(X, T, M, Y)$, we denote by $\theta(F)$ and $\beta(F)$ the ``population" regression coefficients:
\begin{align*}
(\theta(F), \beta(F)) &= \argmin_{(\theta, \beta) \in \R^{\textup{dim}(f) + \textup{dim}(g)}} \E_F \left[ \sum_{\ell = 1}^p [ Y_{\ell} - (\theta, \beta)^{\top}(T f_{\ell}(X), g_{\ell}(X))]^2 \right].
\end{align*}
For any real- or vector-valued function $f \equiv f(x, t, m, y)$ and any $k \in \{ 1, 2 \}$, we denote by $\E_{\D_k}[f(X, T, M, Y)]$ the average of $f(X_i, T_i, M_i, Y_i)$ in the dataset $\D_k$.

\section{Decomposing the observed discrepancy} \label{section:decomposition}

\subsection{A conceptual decomposition}
\label{subsec:concept}

The conceptual foundation for our methods is the following decomposition of the discrepancy in estimated effects:
\begin{align}
& \theta(\D_1) - \theta(\D_2) \nonumber \\
&\quad = \{ \theta(\D_1) - \theta(\D_2) \} - \{ \theta(P) - \theta(Q) \} &\textsf{Sampling variability} \label{eq:sampling_variability}\\
&\quad + \{ \theta(P) - \theta(Q) \} - \{ \theta(P) - \theta(P_X \otimes Q_{Y,T \mid X}) \} &\textsf{Covariate shift} \label{eq:covariate_shift}\\
&\quad + \{ \theta(P) - \theta(P_X \otimes Q_{Y, T \mid X}) \} - \{ \theta(P) - \theta(P_{X, T, M} \otimes Q_{Y \mid X, T, M}) \} &\textsf{Mediation shift} \label{eq:mediation_shift}\\
&\quad + \theta(P) - \theta(P_{X, T, M} \otimes Q_{Y \mid X, T, M}) &\textsf{Residual factors} \label{eq:residual_factors}
\end{align}
Estimates of the components (\ref{eq:sampling_variability}) -- (\ref{eq:residual_factors}) can be used to quantify how much observable distribution shift contributes to the discrepancy $\theta(\D_1) - \theta(\D_2)$.  Social scientists have used conceptually similar decompositions to understand intra-group disparities for over sixty years \citep{kitagawa1955, oaxaca1973wages, blinder1973discrimination}, and more refined ``causal" decompositions are common in mediation analysis \citep{vanderweele2014unification, huber2015pitfalls}.  More recently, \cite{cai2023diagnosing} used a similar decomposition to study performance declines for predictive models.

We now explain why the mathematical expressions in the preceding display merit the intuitive names we have assigned them.

\textsf{Sampling variability} contrasts the observed discrepancy $\theta(\D_1) - \theta(\D_2)$ with the hypothetical discrepancy that we would observe from two \emph{population-wide} experiments, $\theta(P) - \theta(Q)$.   This term includes both purely random variability (e.g. from participant sampling and random treatment assignment) as well as systematic bias from selection in the publication or replication process.  When comparing two large pre-registered experiments, this term will be close to zero.

\textsf{Covariate shift} measures how much the population discrepancy $\theta(P) - \theta(Q)$ would decline if the replication population $Q$ were replaced by an idealized replication population that exactly matches the original covariate distribution while preserving the dependence of the outcome on the covariates, namely $P_X \otimes Q_{Y, T \mid X}$.  If this term is large, it provides evidence that observed covariates capture some important heterogeneity in the effect of $T$ on $Y$.  It also suggests that future replication attempts will be more successful if they are conducted in populations that ``look like" the original study in terms of background characteristics.  On the other hand, if \textsf{Covariate shift} is small, then either $P_X$ and $Q_X$ are already very similar or the observed covariates do not capture much heterogeneity that is relevant  for the treatment effect of interest. 

\textsf{Mediation shift} measures how much further the population discrepancy would reduce if the replication population also matched the original population in terms of the treatment-mediator response.  If this component is large, then it signals that the mediator-outcome relationship is relatively stable but the treatment fails to elicit the same mediation response in the two populations.  On the other hand, if \textsf{Mediation shift} is small, then either $P_{M \mid T, X}$ and $Q_{M \mid T, X}$ are already similar, or $M$ only mediates a small fraction of the total treatment effect.

Finally, \textsf{Residual factors} captures all other differences between $\theta(\D_1)$ and $\theta(\D_2)$, such as those caused by subtle changes in experimental protocol or shifts in the distribution of unobserved moderators.  If this term is large, then it signals that observed distribution shifts cannot account for the difference between the two estimates, and further research or theory is required to understand when we should expect to observe the original effect.

\begin{remark}[Asymmetry]
Some terms in this decomposition are asymmetric in $P$ and $Q$.  This is intentional, since we are mainly interested in understanding whether an ``improved" replication study could have captured the original finding.   It is also statistically necessary, since the overlap conditions in \Cref{assumpton:primitives} may only hold in one direction. 
\end{remark}

\subsection{Estimation without publication bias} \label{section:estimators}

We now explain how we estimate the terms (\ref{eq:sampling_variability}) -- (\ref{eq:residual_factors}) in cases where no publication bias is suspected, e.g. when the original and replication studies are both pre-registered.  Since one goal of this paper is to argue that experimental replications are a useful application area for existing generalizability methods, we do not introduce any methodologically novel estimators here.  As such, we give only high-level descriptions of our approach and leave technical details to the appendix.

To estimate \textsf{Covariate shift}, we re-weight the replication dataset $\D_2$ to match key features of the covariate distribution in the original dataset $\D_1$.  Then, we perform the regression analysis in the re-weighted replication dataset and measure how much the estimated discrepancy declines:
\begin{align*}
\widehat{\textsf{Covariate shift}} = \{ \theta(\D_1) - \theta(\D_2) \} - \{ \theta(\D_1) - \theta(\D_2, \textup{weights} = \hat{w}) \}.
\end{align*}
We estimate $\hat{w}$ using the ``balancing weights" framework for sample standardization \citep{hainmueller2012entropy, zhao2017entropy, josey2022calibration, lee2022improving, lu2023you}.  
Concretely, we find the simplest (lowest entropy) weights $\hat{w}\in \RR^{n_2}$ that make each treatment arm of $\D_2$ match the corresponding treatment arm of $\D_1$ in terms of the means of certain features $\phi : \mathcal{X} \to \R^d$. That is, we minimize the entropy $\sum_{i=1}^{n_2}\hat w_i\log \hat w_i$ while imposing that  $\Etwo[\hat{w} \phi(X)] = \Eone[ \phi(X)]$ and $\Etwo[ \hat{w} T \phi(X)] = \Eone[ T \phi(X)]$, with $\hat w_i\geq 0$ and $\sum_{i=1}^{n_2}\hat w_i=1$.  We typically choose these features $\phi$ so that $\hat{w}$ balances the means (and sometimes variances and covariances) of suspected effect modifiers.

To estimate \textsf{Mediation shift}, we measure how much further the discrepancy declines after we additionally re-weight $\D_2$ to match features of the mediator-treatment relationship in $\D_1$:
\begin{align*}
\widehat{\textsf{Mediation shift}} = \{ \theta(\D_1) - \theta(\D_2, \textup{weights} = \hat{w}) \} - \{ \theta(\D_1) - \theta(\D_2, \textup{weights} = \hat{\omega}) \}.
\end{align*}
We estimate $\hat{\omega}$ using the simplest weights that balance the same covariate moments as $\hat{w}$ and also satisfy $\Etwo[\hat{\omega} \psi(M)] = \Eone[\psi(M)], \Etwo[\hat{\omega} T \psi(M)] = \Eone[T \psi(M)]$ for some mediator features $\psi : \mathcal{M} \to \R^k$.  For continuously-distributed mediators we take $\psi(m) = (m, m^2)$, thereby matching the treatment-mediator correlation in $\D_1$.  For categorical mediators with a handful of levels, we take $\psi(m)$ to be the one-hot encoding of $M$, thereby matching the entire joint distribution of $(T, M)$ in $\D_1$.

To estimate \textsf{Sampling variability}, we use the fact that the observed discrepancy $\theta(\D_1) - \theta(\D_2)$ is an approximately unbiased estimate of the population discrepancy $\theta(P) - \theta(Q)$ when selection bias is absent.  Therefore, we can simply estimate this term by zero.  Finally, we estimate \textsf{Residual factors} by whatever remains:
\begin{align*}
\widehat{\textsf{Sampling variability}} &= 0\\
\widehat{\textsf{Residual factors}} &= \{\theta_1(\D_1) - \theta_1(\D_2) \} - \widehat{\textsf{Covariate shift}} - \widehat{\textsf{Mediation shift}} - \widehat{\textsf{Sampling variability}}.
\end{align*}

To set confidence intervals on each of these components, we use jackknife standard errors along with normal-approximation critical values. 
The properties of our methods, including consistency of point estimation and validity of normal approximation confidence intervals, are detailed in Appendix~\ref{appendix:informal_properties}.

\subsection{Selection bias corrections} \label{section:selection_adjustment}

In the presence of selection bias by journals or replicators, the methods described in \Cref{section:estimators} may have poor statistical properties.  In particular, estimators may be badly biased and confidence intervals may have coverage far below the nominal level \citep{rothstein2005publication, duval2000trim}.  

This section describes how these issues can be addressed using post-selection inference methods, at least under certain strong assumptions on the publication and replication process.  Although the assumptions we impose might not be perfectly realistic, we consider them to be an improvement over simply ignoring selection bias. 

Our first assumption concerns the decision to publish or replicate a study.  Following \cite{hung2020statistical}, we assume that the original study's data influences the selection process only by achieving a ``significant enough" p-value on some regression estimate.

\begin{assumption}[Selection model] \label{assumption:selection}
Let $\hat{\tau}$ denote some regression estimate from the original study, which need not be the target coefficient $\theta(\D_1)$, and let $\hat{\sigma}$ denote its standard error.  Further let $ {p} = 2 \{ 1 - \Phi(| \hat{\tau}/\hat{\sigma}|) \}$ denote original study's (two-sided) p-value for $\hat{\tau}$.  For some significance threshold $\alpha_0 \in (0, 1)$, assume that the original study would have been equally likely to be published and replicated with a different sample as long as $\hat{p}$ remained less than $\alpha_0$:  $\P( \text{Original study published and replicated} \mid \D_1) = \P( \text{Original study published and replicated} \mid  {p} < \alpha_0)$.
\end{assumption}

In our applications, we use the threshold $\alpha_0 = 0.05$.  Smaller thresholds are arguably more realistic but give more conservative results.  This selection model allows the publication and replication process to depend arbitrarily on deterministic study characteristics like research question, study design, or author affiliation.  However, it does not allow the original study authors to engage in specification search (``p-hacking") or for journals to prefer publishing ``very significant" p-values.  

Our second assumption concerns the replication sample size $n_2$.

\begin{assumption}[Replication sample size] \label{assumption:replication_sample_size}
Assume that, conditional on the original study being chosen for publication and replication, the replication sample size $n_2$ is independent of the original data $\mathcal{D}_1$.
\end{assumption}

This assumption is implicitly required to use the methods in \cite{hung2020statistical}, but it is not formally stated there.  We must admit it is rather unrealistic, since replication studies typically choose their sample sizes to achieve a certain power against the original study's point estimate.  Nevertheless, \Cref{appendix:simulations} shows via extensive simulations that the selective inference methods described below remain robust to such violations of \Cref{assumption:replication_sample_size} in practical situations.

Let $\widehat{\textsf{Component}}$ denote any of the estimators from \Cref{section:estimators}.  Under Assumptions \ref{assumption:selection} and \ref{assumption:replication_sample_size}, the theory of post-selection inference implies that the conditional-on-replication distribution of $\widehat{\textsf{Component}}$ is approximately truncated normal \citep{hung2020statistical}.  Thus, if we let $\hat{\Sigma}$ denote the jackknife estimate of $\Cov\{ \theta(\D_1), \widehat{\textsf{Component}} \}$, then results in 
\cite{lee2016exact, kasy2019identification} imply that the following confidence interval has approximately $100(1 - \alpha)\%$ coverage even conditional on the original study being published and replicated:
\begin{align}
\begin{split}
&\text{CI}(\alpha) = \left\{ t \, : \, p(t, \widehat{\textsf{Component}}, \theta(\D_1), z_{1 - \alpha/2} \hat{\sigma}_0, \hat{\Sigma}) \in [\alpha/2, 1 - \alpha/2] \right\}\\
&\quad \text{where } p(t, c, \vartheta, \tau, \Sigma) := \P_{Z \sim N(t, \Sigma_{11})} \{ Z \leq t \mid \tau < | \Sigma_{11}^{-1} \Sigma_{12}(Z - c) + \vartheta| \}
\end{split}
\end{align}
For point estimation, we maximize the truncated Gaussian likelihood.  See Appendix~\ref{appendix:selective_mle} for implementation details of selection-adjusted point estimation and confidence intervals.

\subsection{Building intuition with stylized examples} \label{section:stylized_examples}

We now build some intuition about how to read and interpret our proposed discrepancy decomposition by applying it in a number of stylized simulated examples with a unambiguous ``correct" answers.  For simplicity, we will only focus on pre-registered experiments in these stylized examples and apply the methods of \Cref{section:estimators}.

\begin{figure}[htbp]
    \centering
    \begin{subfigure}[b]{0.49\textwidth} \centering
       \includegraphics[width=8.4cm]{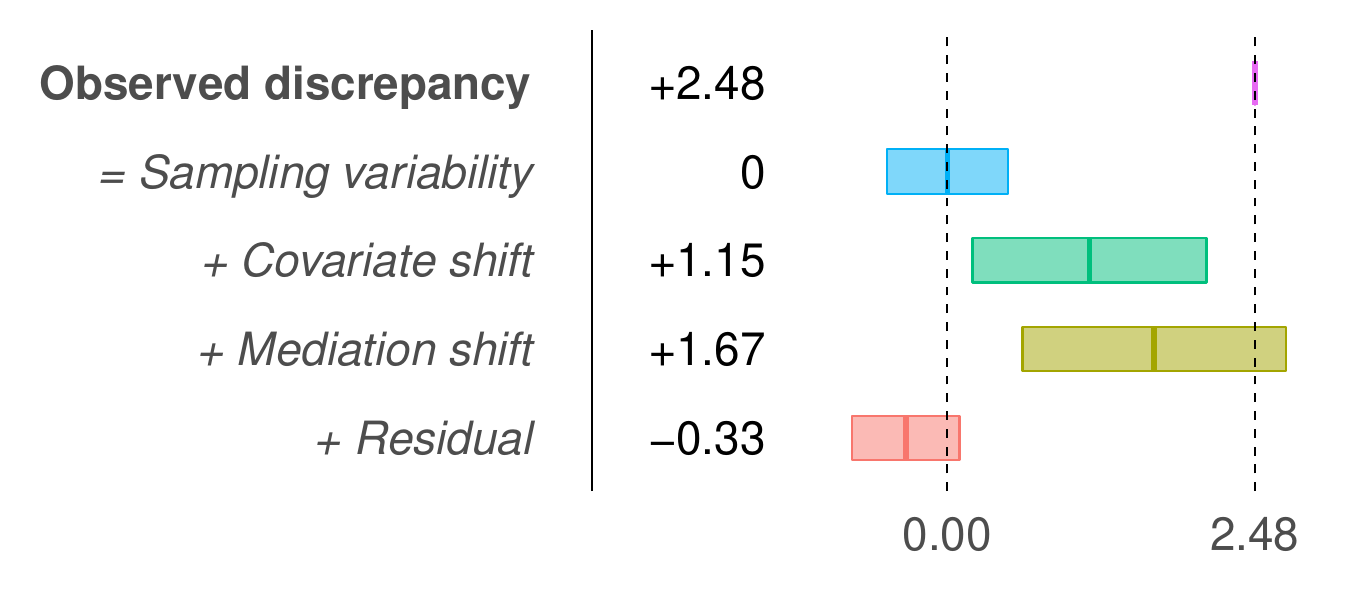} 
       \caption{Decomposition for \Cref{example:observed_heterogeneity}.  The estimated components corresponding to \textsf{Covariate shift} and \textsf{Mediation shift} are large and significantly different from zero.} \label{fig:observed_heterogeneity_dominates}
    \end{subfigure}
    \hfill
    \begin{subfigure}[b]{0.49\textwidth}
    \includegraphics[width=8.4cm]{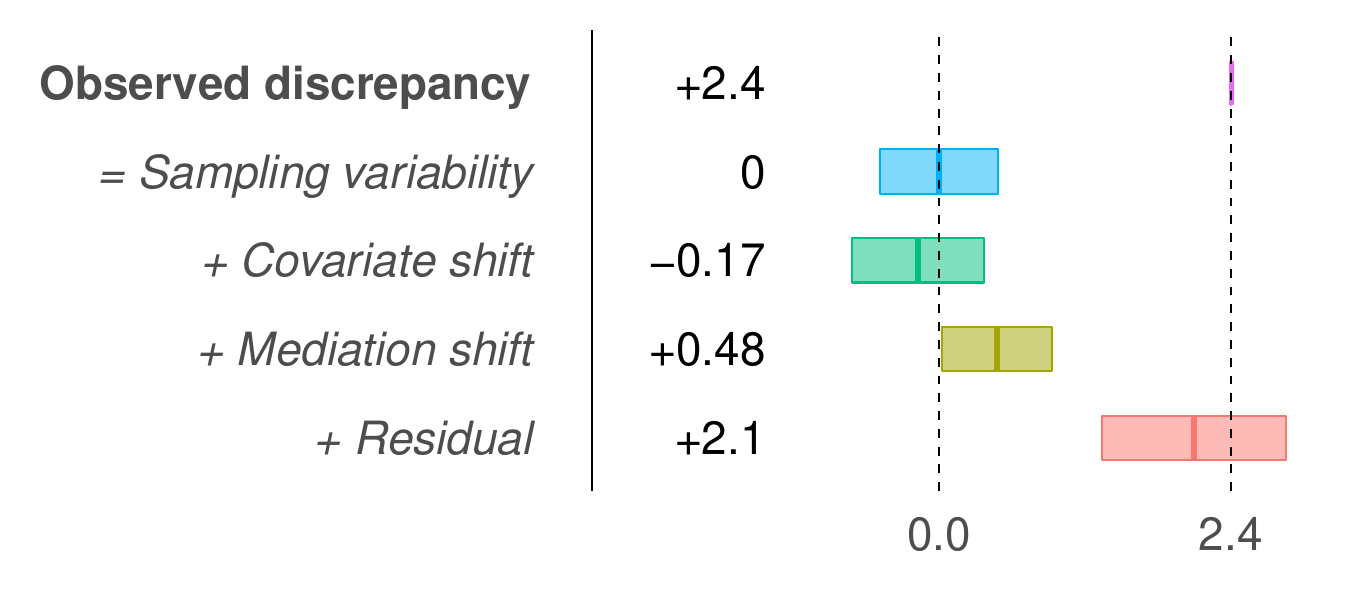}
    \caption{Decomposition for \Cref{example:hidden_moderator}.  The estimated \textsf{Residual} component is large and significantly different from zero.  This is due to distribution shift on a key hidden moderator.}
    \label{fig:hidden_moderator}
    \end{subfigure}
    \caption{Decomposition for the stylized examples in \Cref{section:stylized_examples}.  Shaded bars are 90\% confidence intervals.   In Panel (a), the decomposition suggests that a better-implemented replication study conducted in a population more similar to $P$ on observed background characteristics would estimate an effect closer to the original study's.  In Panel (b), an improved replication study would also need to match the distribution of the hidden moderator $U$ to reduce the discrepancy from the original study.
    }
\end{figure}

\begin{example}[Shift in observable characteristics] \label{example:observed_heterogeneity}
Our first example shows what our decomposition results look like when observable distribution shifts can explain the discrepancy.

We consider a hypothetical educational intervention $T$ that encourages students to do additional background reading before class.  Only a single covariate $X$ is present (which we will call ``Age") and a mediator $M \in \{ 0, 1 \}$ records whether participants indeed engaged in extra reading.  The outcome $Y$ is a continuous measure that can be thought of as a test score.

In the original population, the participants are freshman- or sophomore-aged college students and the treatment increases the chance of doing background reading by 50\%.  However, the replication study samples a broader student population and implements the treatment poorly, so the chance of doing additional reading increases by only 25\%.
\begin{align*}
P_X &= N(19, 0.5^2) \quad \text{and} \quad P_{M \mid X, T} = \text{Bernoulli}(0.1 + 0.50 T)\\
Q_X &= N(21, 1.5^2) \quad \text{and} \quad Q_{M \mid X, T} = \text{Bernoulli}(0.1 + 0.25 T)
\end{align*}
The two populations are otherwise identical, with $P_{Y \mid X, T, M} = Q_{Y \mid X, T, M} = N( X + 2M \{ 22 - X \}, 1)$.  We randomly sampled $n_1 = n_2 = 500$ participants from these populations to form the datasets $\D_1$ and $\D_2$, and assigned treatments $T$ by a fair coin flip for each participant.

In the original dataset, an OLS regression of $Y$ on $T$ and $X$ estimated an effect of $\approx 3.16$.  However, the same analysis in the replication dataset yielded a much smaller effect size of $\approx 0.68$.  To diagnose the large discrepancy between these two estimates, we applied \Cref{section:estimators}'s methods.  To adjust for covariate shift, we re-weighted each treatment arm of $\D_2$ to match the mean and variance of ``Age" in the corresponding treatment arm of $\D_1$.  To adjust for mediation shift, we additionally matched the $(T, M)$ joint distribution.  

Decomposition results are displayed in \Cref{fig:observed_heterogeneity_dominates}.  The first thing to notice is that the 90\% confidence interval around \textsf{Sampling variability} does not reach the observed discrepancy $ \theta(\mathcal{D}_1) -  \theta(\mathcal{D}_2) = 3.16-0.68  = 2.48$, indicating that sampling variability alone does not explain the observed discrepancy.  However, the \textsf{Residual} component is small and statistically insignificant, suggesting that observed heterogeneity \emph{can} explain away the observed discrepancy.  More concretely, about half the observed discrepancy can be attributed to each of \textsf{Covariate shift} and \textsf{Mediation shift}.

A decomposition like Example~\ref{example:observed_heterogeneity} is fairly actionable.  It suggests that future replicators can reasonably expect to reproduce the original finding if they sample an appropriately-aged population and implement the treatment with enough care. ${\Box}$
\end{example}

\begin{example}[Shift in hidden moderator] \label{example:hidden_moderator}
Our second example shows that decomposition results look like when the distribution of some hidden moderators changes across studies. 

We consider the same experiments as \Cref{example:observed_heterogeneity} but with a different outcome variable, which we also think of as a test score.  With a slight abuse of notation, we will also call this outcome ``$Y$".

For this new outcome, we assume that the effect of background reading ($T = 1$) is moderated by an unobserved factor $U \in \{ 0, 1 \}$ which measures whether a student has taken an introductory calculus class.  In both populations, the outcome is sampled as $Y \mid X, T, M, U \sim N(10 + M\{ 1 + 5U \}, 1)$, meaning that extra reading is much more effective for students who know calculus.

We also introduce distribution shift on the unobserved moderator.  In the original study's university, calculus is a required class so $P(U = 1) = 1$.  However, the replication study is conducted in a university where calculus is optional and $Q(U = 1 \mid X) = 1/\{ 1 + \exp(21-X) \}$.  This means about half of the replication study population has taken calculus, but most of these are older students.

Using the same samples from \Cref{example:observed_heterogeneity}, we estimated treatment effects on this secondary outcome by regressing $Y$ on $T$ and $X$ with OLS.  This yielded an effect size of $\approx 3.05$ in the original dataset and $\approx 0.65$ in the replication dataset.  Thus, we once again find the effect size declines substantially across studies.

To diagnose the discrepancy, we performed our decomposition analysis with the same weights in \Cref{example:observed_heterogeneity}.  The results are displayed in \Cref{fig:hidden_moderator}.  In this decomposition, the first thing to notice is that the \textsf{Residual} component is large and 
statistically significant.  This leads us to conclude that observable distribution shift \emph{cannot} explain the non-replication (although there is reasonable evidence for some \textsf{Mediation shift}).  

An interesting feature of this example is that the true value of \textsf{Covariate shift} is negative.  This means that conducting the replication study on students more similar to the original study in terms of ``Age" would actually \emph{increase} the expected discrepancy.  This is because decreasing the ``Age" discrepancy increases the ``Calculus" discrepancy.  When hidden moderators are present, we should not always expect correcting observed distribution shifts to bring point estimates closer together. ${\Box}$
\end{example}

\section{Real-world examples} \label{section:case_studies} 

We now apply our discrepancy decomposition to data from three directly-replicated experiments in behavioral science (clinical psychology, social psychology and experimental economics).  The reason we focus on behavioral science experiments is due to data availability:  in our experience, this community has made the strongest push to conduct direct replications and make experimental data publicly available. Analysis for other data pairs (with arguably less rich features) can be found in our data repository.\footnote{See \url{https://github.com/ying531/awesome-replicability-data} for the analysis and cleaned datasets.}

None of the original experiments analyzed in this section were pre-registered.  As a result, there may be some degree of selection bias present.  To address this bias, we report selection-adjusted decompositions using the methods of \Cref{section:selection_adjustment} with a significance threshold of $\alpha_0 = 0.05$ alongside -- or in place of -- the basic methods from \Cref{section:estimators}. 

\subsection{Eye movement and false memory} \label{section:emdr}

Our first case study looks at two clinical psychology studies on the relationship between eye movement and false memory.  We will see that the replicators' conjectured covariate shift does not explain away the discrepancy between the two study's estimates.

Both studies investigate whether engaging in repeated lateral eye movement while thinking about a traumatic event enhances susceptibility to misinformation concerning that event.  Such eye movements are a standard component of ``eye movement desensitization and reprocessing" therapy for posttraumatic stress disorder.

The original study by \cite{houben2018lateral} recruited $n_1 = 82$ undergraduate students from Maastricht University in the Netherlands.  Each participant was instructed to carefully watch a video of a traumatic car crash.  Then, half of the participants were assigned to think about the video while watching a grey dot move laterally on a black screen (the ``treatment").  The remaining participants were assigned to think about the video while watching a stationary dot (the ``control").  Afterwards, each participant read a misinformation-laden narrative from an eyewitness and answered 15 forced-choice questions concerning details of the crash.

The replication study by \cite{calvillo2019lateral} recruited $n_2 = 120$ undergraduate students from California State University San Marcos in the United States.  The experimental protocol was largely the same as the original study.  However, after each forced-choice question, the experimenters also asked participants how they selected their answers.

The two studies came to rather different conclusions.  While the original study found that treated participants correctly answered $\approx 1.14$ fewer questions on average than control participants ($t$-test p-value $< 0.001$) and endorsed $0.85$ additional pieces of misinformation ($t$-test p-value $< 0.001$).  However, the replication study found that treated participants answered only $0.09$ fewer questions correctly ($t$-test p-value $\approx 0.68$) and endorsed $-0.03$ additional pieces of misinformation ($t$-test p-value $\approx 0.83$).  

What is responsible for the large discrepancy between the two studies' estimates?  One conjecture advanced by Calvillo et al. was covariate shift:
\begin{quote}
``\textit{We noticed in our exploratory analyses ... that the depression scores for participants in the present study were considerably higher than those in Houben et al.'s (2018) study ... The differences between the two samples may have contributed to the failure to replicate the primary findings.}"
\end{quote}
Indeed, over 13\% of participants in the replication study had ``moderate" or ``severe" depression ratings on Beck's Depression Inventory (BDI) while all participants in the original study had ``mild" or ``minimal" depressive scores\footnote{The original study evaluated depression level using BDI-I and the replication study used BDI-II.  To make these comparable, we discretized these into categorical depression ratings using the standard cutoffs (see \cite{beck1988psychometric} for BDI-I and \cite{smarr2011measures} for BDI-II).}  There are also some plausible mediation shifts.  Eye movements are believed to influence recall by reducing the vividness and emotionality associated with traumatic memory.  In the replication study, the estimated declines in self-reported ``vividness" and ``emotionality" caused by the treatment were larger than in the original study, although not significantly so.

To evaluate these conjectures, we applied our discrepancy decomposition on each of the two outcomes considered.  We estimated the impact of \textsf{Covariate shift} by re-weighting each treatment arm of $\D_2$ to match the corresponding treatment arm of $\D_1$ on the following factors:
\begin{itemize}[itemsep=-1ex]
\item The proportion of students with ``minimal", ``mild", ``moderate" and ``severe" depression ratings.
\item The average age.
\item The percent of female participants.
\item The average ``vividness" and ``emotionality" scores assigned at baseline (before the treatment).
\end{itemize}
To estimate the impact of \textsf{Mediation shift}, we further matched the mean and variance of the post-treatment ``vividness" and ``emotionality" scores in each treatment arm.  This makes the treatment-mediator correlations in $\D_2$ agree with those in $\D_1$.   

\begin{figure} 
    \centering
    \begin{subfigure}[b]{0.49\textwidth} \centering
       \includegraphics[width=8.4cm]{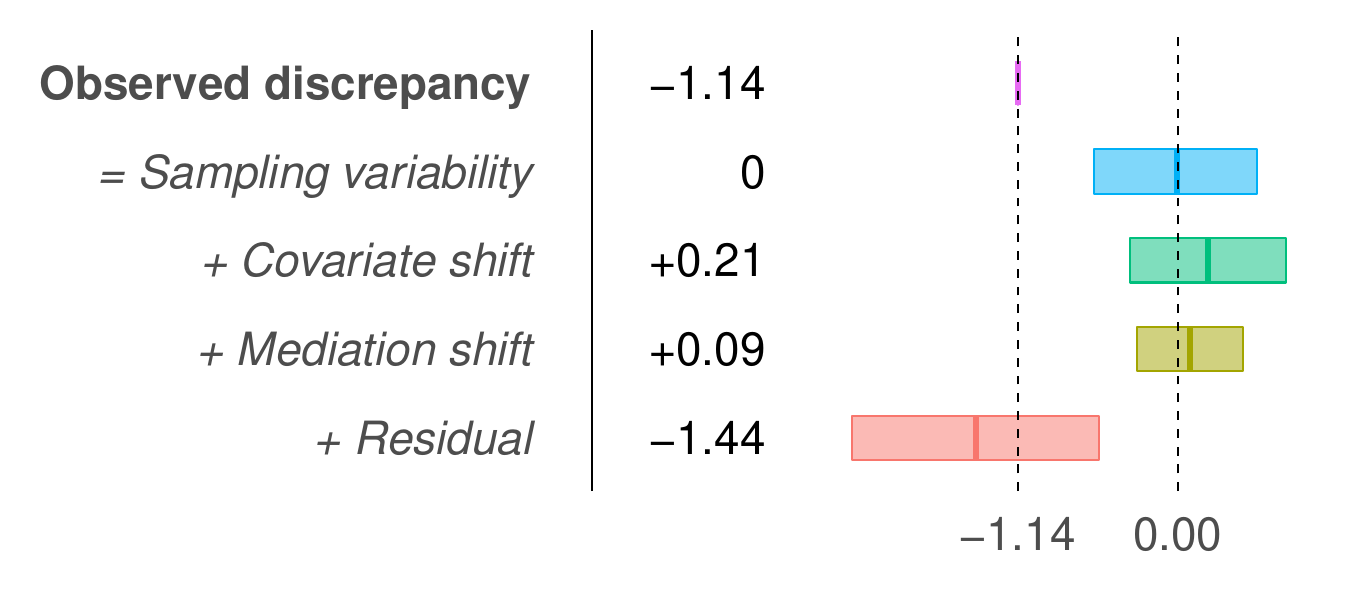} 
       \caption{Decomposition for ``Total correct answers"} \label{fig:emdr_totalcorrect}
    \end{subfigure}
    \hfill
    \begin{subfigure}[b]{0.49\textwidth}
    \includegraphics[width=8.4cm]{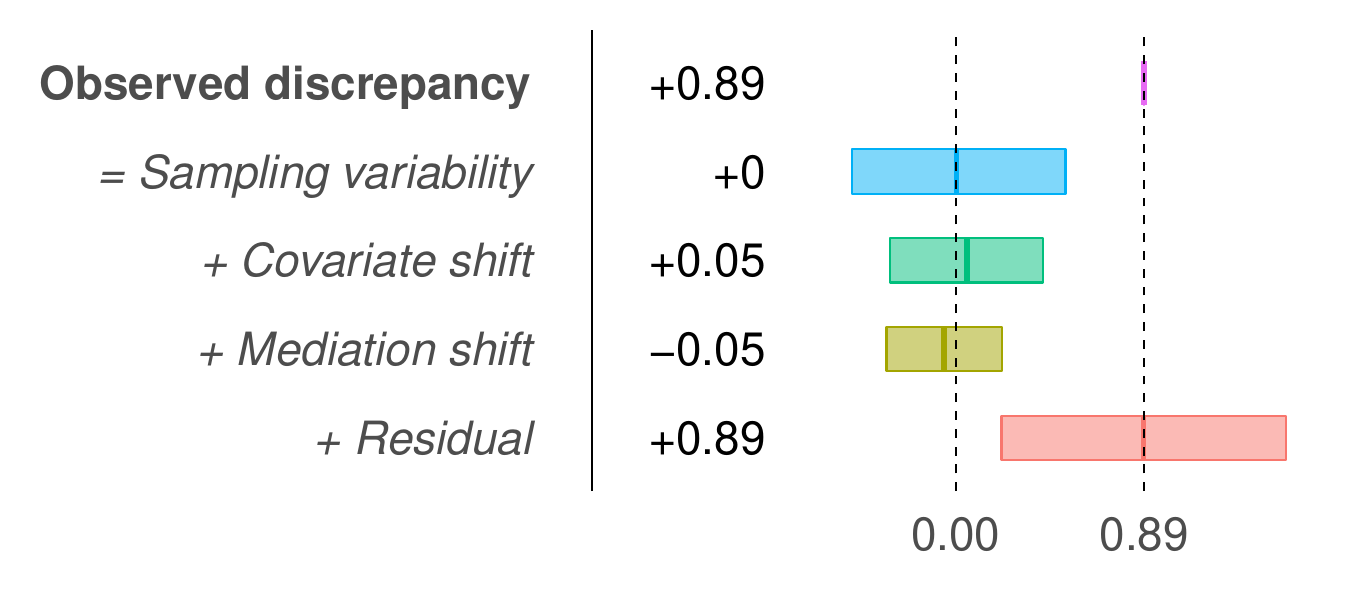}
    \caption{Decomposition for ``Misinformation endorsed"}
    \label{fig:emdr_misinfo}
    \end{subfigure}
    \caption{Decompositions for the two lateral eye movement experiments described in \Cref{section:emdr}, with panels (a) and (b) studying different outcome variables.  The shaded bars are 90\% confidence intervals.  Point estimates and confidence intervals are corrected to account for selection bias attributable to the fact that the original study's point estimates are statistically significant at the $\alpha_0 = 0.05$ level.}
    \label{fig:emdr_decomposition}
\end{figure}

Decomposition results are displayed in \Cref{fig:emdr_decomposition}.  The first thing to notice is that the 90\% confidence intervals around the \textsf{Sampling variability} components do not reach the observed discrepancies, so random factors -- including selection bias of the form assumed in \Cref{assumption:selection} -- cannot fully explain the observed non-replication.  Indeed, the p-values achieved in the original experiment were so small that our selection model implies essentially no systematic sampling bias.  One caveat is that our confidence intervals assume independent sampling of participants, while the actual participants in these experiments were volunteers.

More surprisingly, both \textsf{Residual} components are large and statistically significant.  This suggests that observable distribution shift --- including the covariate shift hypothesized by the replicators --- does \emph{not} convincingly explain the discrepancy in effect estimates.  In fact, we estimate that correcting for \textsf{Covariate shift} and \textsf{Mediation shift} would make little (if any) difference in reconciling the two study results.

Our takeaway from this analysis is that some as-of-yet unidentified implementation detail or moderating factor differs across the two studies and drives the concept shift $P_{Y \mid X, T, M} \neq Q_{Y \mid X, T, M}$.  While cultural differences between the Netherlands and the United States may seem like a natural candidate, we will mention that a second direct replication by \cite{van2019lateral} in the Netherlands also estimated little-to-no effect for this treatment.  The task of identifying the unobserved factors responsible for this non-replication goes beyond data analysis and requires additional psychological theory.

\subsection{Emotion and time preference} \label{section:emotion_time_preference}

Our next case study looks at two experimental economics experiments on the impact of mild positive emotion on time preference.  The second experiment was conducted as part of the Experimental Economics Replication Project \citep{camerer2016evaluating}.  In this case, we will find that publication bias or unobserved factors are more compelling discrepancy explanations than the conjectured mediation shift.

The original study by \cite{ifcher2011happiness} recruited $n_1 = 69$ undergraduate students from Santa Clara University in the United States.  Of these, 34 were randomly assigned to watch a positive-affect-incuding comedy montage (the ``treatment").  The remaining 35 students watched a neutral-affect-inducing nature clip (the ``control").  Subsequently, each student gave their subjective valuations of 30 delayed-payment contracts.  Thus, the outcome in this experiment is a length-30 vector $Y_i = (Y_{i,1}, \dots, Y_{i,30})$.  Although several different analyses were performed in the original study, we will focus on the two-way ANOVA analysis reported in \citet[Table 3 Column 5]{ifcher2011happiness}.

The replication study by \cite{camerer2016happinessreplication} recruited $n_2 = 168$ students from Nuffield College in the United Kingdom.  Of these, 86 watched the comedy montage clip from the original study, and the remaining 72 watched the nature clip.  As in the original study, each student then valued 30 delayed-payment contracts.

While the original study estimated that the treatment \emph{increased} present values by approximately 3 dollars on average ($p = 0.027$), the replication estimated that the treatment \emph{decreased} present values by approximately 6 cents on average ($p = 0.966$).

In their replication report, Camerer et al. mainly attributed this discrepancy to mediation shift.  In particular, they argued that the comedy montage treatment did not elicit the same emotional response in their study as in the original study, perhaps because of cultural differences between American and British audiences, comedic preferences changing in the five-year changing between the two studies, or the well-known death of the comedian featured in the ``treatment" condition's montage.  They write:
\begin{quote}
``\textit{Mood induction is the key in this experiment.  It had the intended effect on affect in the original article ... In the replication study, mood inducement did not have the intended effect on affect."}
\end{quote}
As evidence, the replicators reported that the average ``PANAS net positive affect" score (a measure of mood, with higher values indicating better moods) was actually lower in their treatment group than in their control group, reversing the trend from the original study.  The proportion of treated students who reported that the clip made them feel ``sadder" was also higher in the replication study ($13\%$, vs. $6\%$ in the original study).

We applied our discrepancy decomposition to this data to quantify the role played by mediation shift and other observed factors.  Although covariate shift was not a primary focus for the replicators, we nevertheless accounted for some large demographic shifts by re-weighting each treatment arm of $\D_2$ to match the corresponding treatment arm of $\D_1$ in terms of the proportion of Christian students, religiously practicing students, and white students.  To estimate the impact of \textsf{Mediation shift}, we further matched these treatment arms on the following dimensions:
\begin{itemize}[itemsep=-1ex]
\item The proportion of students who reported that the clip made them feel ``happier", ``sadder" and ``neither happier nor sadder".
\item The proportion of students who reported that the clip put them in ``a better mood", ``a worse mood" and ``neither a better nor a worse mood".
\item The mean and variance of the ``PANAS net positive affect" mood indicator.
\end{itemize}

\begin{figure} 
    \centering
    \begin{subfigure}[b]{0.49\textwidth} \centering
       \includegraphics[width=8.4cm]{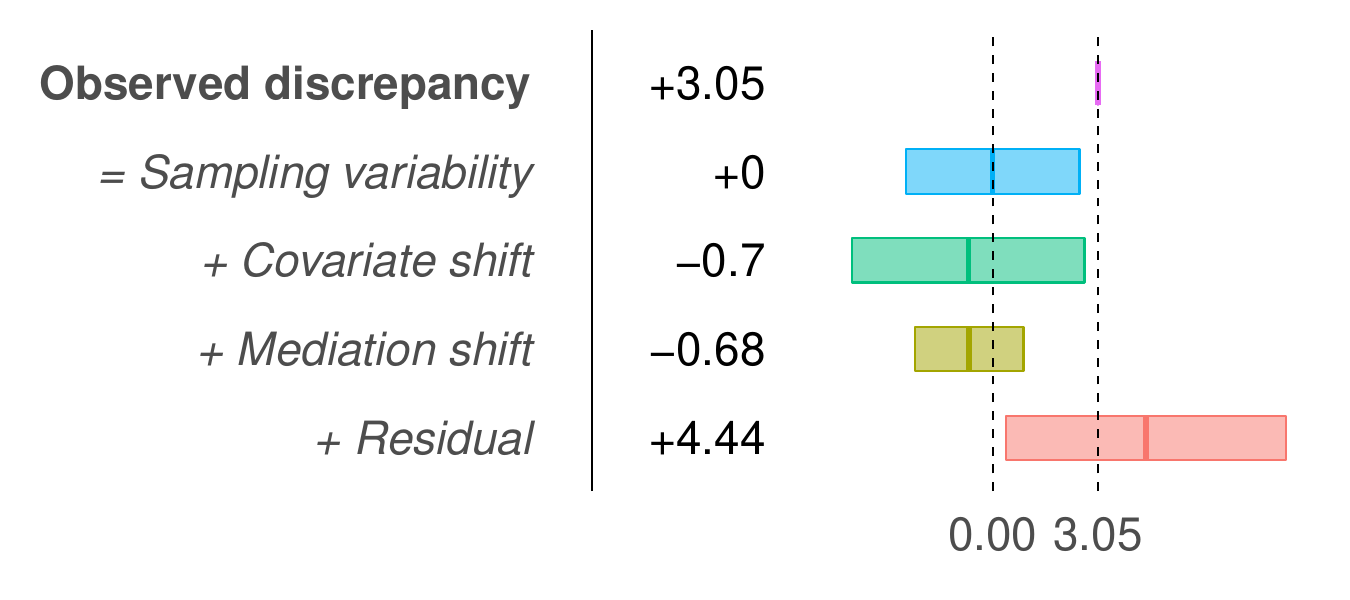} 
       \caption{No selection bias correction} \label{fig:economics_noselection}
    \end{subfigure}
    \hfill
    \begin{subfigure}[b]{0.49\textwidth}
    \includegraphics[width=8.4cm]{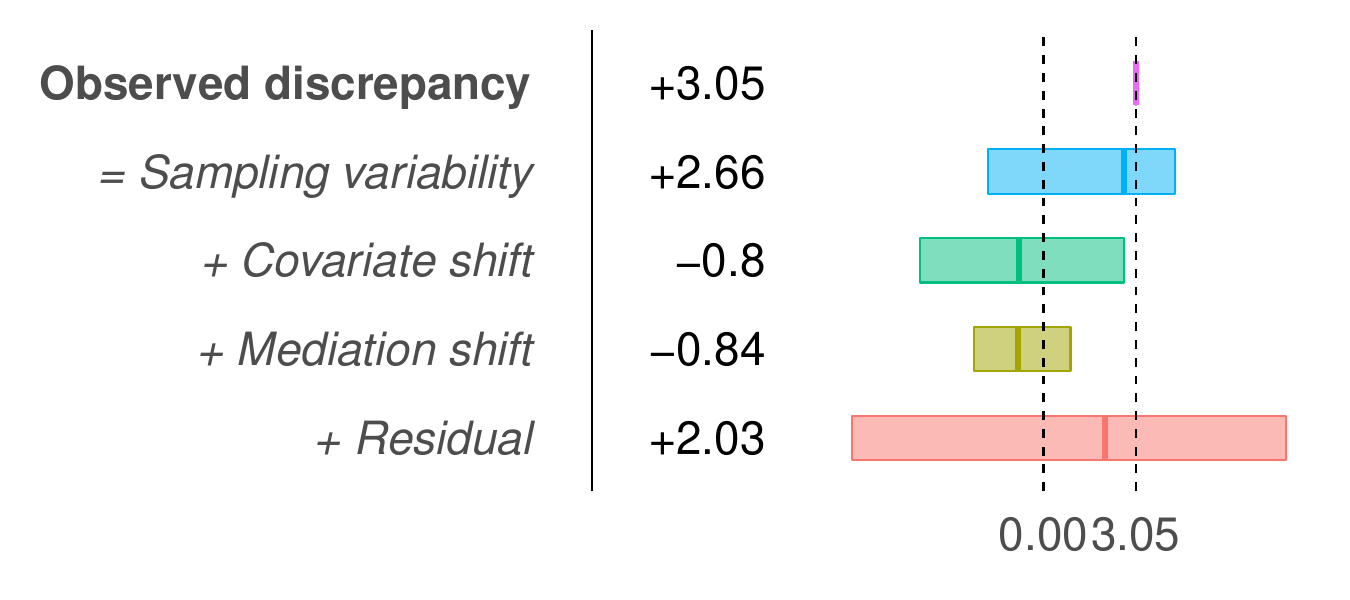}
    \caption{Selection bias correction with $\alpha_0 = 0.05$}
    \label{fig:economics_noselectoin}
    \end{subfigure}
    \caption{Decompositions for the two time preference experiments described in \Cref{section:emotion_time_preference}.  The shaded bars are 90\% confidence intervals.  In panel (a), we assume no publication bias is present and use the methods of \Cref{section:estimators}.  In panel (b), we account for publication bias attributable to the original estimate's p-value falling below $\alpha_0 = 0.05$.}
    \label{fig:economics_decomposition}
\end{figure}

We performed one decomposition that assumes no publication bias is present and one that accounts for the original estimate's p-value falling below $0.05$.  The results are visualized in \Cref{fig:economics_decomposition}.  

One surprising feature is common across the two analyses:  neither decomposition estimates that accounting for \textsf{Covariate shift} or \textsf{Mediation shift} would reduce the effect size discrepancy.  Thus, a two-way ANOVA model estimated in a weighted version of $\D_2$ which matches $\D_1$ on the treatment-PANAS correlation and the distribution of self-reported ``happiness" status in each treatment arm (among other things) would still estimate a negative average treatment effect.

The takeaway from the two analyses is otherwise quite different.  Under a ``no publication bias" assumption (i.e. the original study would have been equally likely to be published with any p-value), then the large and statistically significant \textsf{Residual} component in \Cref{fig:economics_noselection} suggests that the difference between the two estimates is driven by some unobserved factor.  For example, an unmeasured component of mood not captured by self-reported happiness/mood might contribute to this \textsf{Residual}.  Under the weaker assumption that the original study would have been equally likely to be published with any p-value less than $0.05$, the \textsf{Residual} component is no longer significant.  Instead, publication bias and sampling variability alone could entirely explain the observed discrepancy.

\subsection{Accuracy nudge and Covid-related misinformation} \label{section:covid_misinfo}

Our last example looks at two psychology experiments investigating the effect of a nudge intervention on individuals' decision to share fake news regarding COVID-19.  The second experiment was conducted as part of the Center for Open Science's project on ``Systematizing Confidence in Open Research and Evidence". 
What is special about these studies is that they both collected covariates that are highly relevant to the phenomenon of interest. In this example, the signal-to-noise ratio is too low to draw clear conclusions. That being said, our analysis suggests that adjusting for covariate shifts explains some of the discrepancy for the main effects, but not for the original study's interaction coefficient. 

The original study by \cite{pennycook2020fighting} was conducted on a sample of $n_1 = 856$ participants\footnote{Pennycook et al. conducted two experiments.  The replication focused on Study 2.  The reported sample size excludes participants who failed to complete the survey} from the online platform Lucid, which uses quota sampling to match the general U.S. population on various demographic features.  Each participant was randomly assigned to one of two conditions.  Participants in the ``treatment" condition first rated the accuracy of a politically-neutral headline unrelated to COVID-19 and then rated how likely they were to share each of 30 COVID-related headlines.  Among these headlines, 15 were fake news.  Participants in the ``control" condition directly rated the COVID-related headlines without first rating the accuracy of an unrelated neutral headline.

The replication study by \cite{roozenbeek2021accurate} recruited $n_2 = 1,583$\footnote{The participants in Roozenbeek et al. were recruited in an adaptive two-stage sampling plan.  As in \cite{roozenbeek2021accurate}, we do not account for this adaptivity in our analysis.} participants through the online platform Respondi, which also used quota sampling.  The experimental protocol was largely the same as in the original study.  The main difference is that 30 new COVID-related headlines were used, as the original headlines had lost relevance in the period between the two experiments.

Using a two-way ANOVA model with ``treatment" and ``headline veracity" factors, the original study authors estimated a substantial interaction between the two factors ($\hat{\beta} \approx 0.0343, p < 0.001$).  Thus, the authors concluded that their accuracy nudge intervention had a positive effect on ``discernment."  The replication study also estimated a significant interaction effect, but at a much smaller magnitude ($\hat{\beta} \approx 0.015, p \approx 0.017$).  Moreover, the \emph{mechanism} driving the interaction coefficient differed in the two studies:
\begin{itemize}[itemsep=-1ex]
    \item In the original study, the treatment increased participants' willingness to share \emph{true} headlines, but had no detectable impact on their (un)willingness to share false headlines.
    \item In the replication study, the treatment decreased participants' willingness to share \emph{false} headlines, but had no detctable impact on their williness to share true headlines.
\end{itemize}

In our analysis, we investigate whether \textsf{Sampling variability} and \textsf{Covariate shift} are able to explain these discrepancies.  No mediators were measured in the original study, so we omit the \textsf{Mediation shift} component from our decomposition.

Despite the use of quota sampling, the two studies differed substantially on many study-relevant background characteristics.  For example, participants in the replication study scored 20\% higher on average in a Cognitive Reflecion Test and 10\% higher on average in a scientific knowledge quiz than participants in the original study ($p < 0.0001$ for both comparisons).  Participants in the original study were also more likely to lean Republican than participants in the replication study.

To quantify the contributions of observed covariate shifts to the effect-size discrepancy, we re-performed the replication analysis in a weighted version of $\mathcal{D}_2$ that matches each treatment arm of the original dataset on the following factors\footnote{In the original dataset, scientific knowledge scores were mean-imputed for 2 participants and political leanings were mode-imputed for 5 participants.}:
\begin{itemize}[itemsep=-1ex]
    \item The mean and variance of the Cognitive Reflection Test (CRT) score.
    \item The mean and variance of the scientific knowledge quiz score.
    \item The mean and variance of the Medical Maximizer-Minimizer Scale (MMS), which is a test measuring the extent to which people seek healthcare for minor issues (``medical maximizers") or eschew non-essential healthcare (``medical minimizers").
    \item The distribution of self-reported political leanings (``Strongly Republican", "Republican", ``Lean Republican", ``Lean Democrat", ``Democrat", ``Strongly Democrat").
    \item The gender distribution.
    \item The mean and variance of age.
\end{itemize}

\begin{figure}
    \centering
    \includegraphics[width=8.4cm]{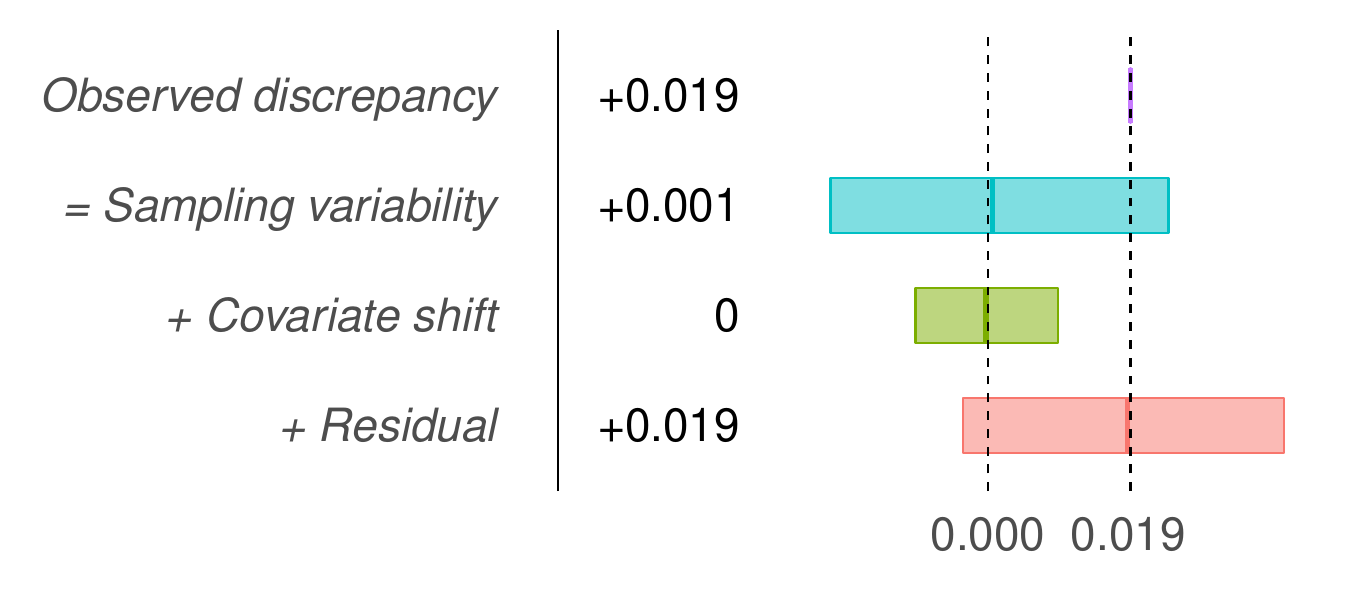}
    \caption{Discrepancy decomposition for the interaction of ``headline veracity" and ``treatment" in the COVID-related misinformation experiments described in \Cref{section:covid_misinfo}.  The shaded bars are 90\% confidence intervals.  The decomposition adjusts for selection bias attributable to the original estimate's p-value falling below $\alpha_0 = 0.05$.}
    \label{fig:covid_interaction}
\end{figure}

\Cref{fig:covid_interaction} shows the results of our decomposition analysis for the interaction coefficient on ``treatment" and ``headline veracity."  The first thing to notice is that the confidence interval on \textsf{Sampling variability} includes the observed discrepancy.  Thus, the difference in observed regression coefficients could plausibly be explained by statistical noise.  Secondly, the point estimate on \textsf{Covariate shift} is too small to even be displayed.  In other words, even though the two study populations differ significantly on several (apparently) study-relevant demographic variables, essentially none of the observed discrepancy is explained by these differences.  This is consistent with Pennycook et al.'s moderation analyses, which found:
\begin{quote}
``\textit{[T]he treatment effect on sharing discernment was not significantly moderated by CRT performance, science knowledge, partisanship ... or MMS score}."
\end{quote}
Roozenbeek et al. also did not attribute any discrepancy to these demographic differences.  Instead, they pointed to factors such as shifts in COVID attitudes over time as possible explanations.

\begin{figure} 
    \centering
    \begin{subfigure}[b]{0.49\textwidth} \centering
       \includegraphics[width=8.4cm]{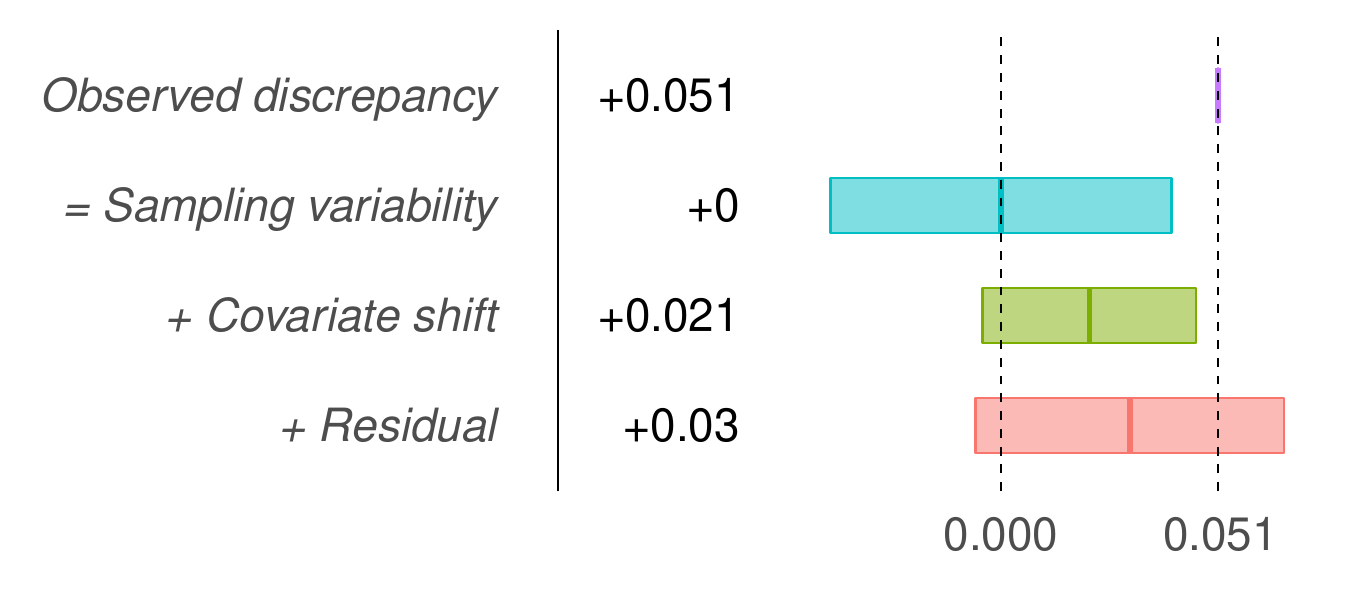} 
       \caption{Decomposition for true headlines} \label{fig:covid_true}
    \end{subfigure}
    \hfill
    \begin{subfigure}[b]{0.49\textwidth}
    \includegraphics[width=8.4cm]{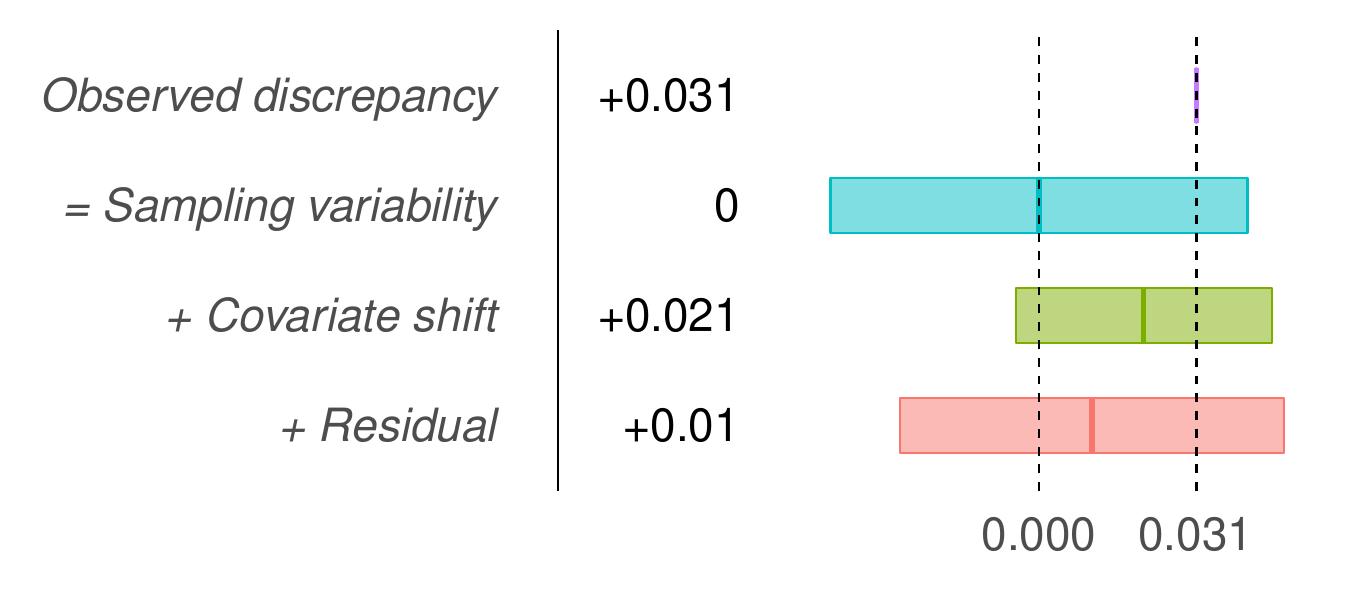}
    \caption{Decomposition for false headlines}
    \label{fig:covid_false}
    \end{subfigure}
    \caption{Discrepancy decompositions for the average treatment effect on ``true headlines" and ``false headlines" in the COVID-related misinformation experiments described in \Cref{section:covid_misinfo}  The shaded bars are 90\% confidence intervals.  Point estimates and confidence intervals are corrected to account for selection bias attributable to the fact that the original study's interaction coefficient on ``treatment" and ``headline veracity" achieved a p-value less than $\alpha_0 = 0.05$.} \label{fig:covid_ttests}
\end{figure}

However, \Cref{fig:covid_ttests} suggests that the preceding analyses may have been too quick to dismiss the role of observed covariates like CRT score.  The left and right panels of this figure show our decomposition results for the $t$-test analyses on ``true" and ``false" headlines, respectively.  In both panels, we estimate that accounting for observed distribution shifts would roughly cut the discrepancies in half.  In particular, the average treatment effect on willingness to share \emph{true} headlines would increase and the average effect on \emph{un}willingness to share \emph{false} headlines would decrease.  However, since the interaction coefficient analyzed in \Cref{fig:covid_interaction} is the difference of these two effects, these two discrepancy reductions cancel each other out and the role of covariates thus goes unnoticed.  We must caveat that the \textsf{Covariate shift} estimates are not statistically significant in either case, so these interpretations should be regarded as merely suggestive.

\section{Discussion}

This study investigates how much of the effect discrepancy between an original study and its replication is attributable to sampling uncertainty, covariate shift, mediation shift, and residual factors. Ideally, discrepancies between studies correspond to changes in observed characteristics, since in that case we can explain why a result did not replicate and predict whether a statistical result will still hold in a new situation. 

In several empirical examples, we found that the contribution of covariate shift and mediation shift was relatively small. On the other hand, residual shifts sometimes explained a large part of the observed discrepancy. Residual shifts capture shifts in unobserved variables, such as subtle changes in treatment implementations, changes in how the outcome is measured, and other shifts in unobserved moderators. This is somewhat concerning since this means that we often do not understand how far a result will generalize.

There are several potential ways to deal with this issue. First, we can collect additional covariates and mediators. Intuitively, as we capture more moderators and mediating variables, residual shifts should get smaller. However, this potential solution is hard to implement in practice since it is not clear how we can make sure that all moderators are measured.

Secondly, instead of thinking about residual distributional changes as \emph{bias}, we can think about it as \emph{variance} that arises through (unintended) changes in the data collection process. This implies that for uncertainty quantification we should not only account for sampling variability but also random distributional shifts. \cite{jeong2022calibrated} discuss statistical inference under random distribution shifts, based on a symmetry assumption.
The point that standard statistical procedures might be inappropriate for uncertainty quantification have been made throughout the statistical literature. For example, \cite{meng2018statistical} introduces a data defect index that measures the representativeness of a sample for a population.
\cite{cox2015big} writes ``big data are likely to have complex structure, in particular implying that estimates of precision obtained by applying standard statistical procedures are likely to be misleading, even if the point estimates of parameters themselves may be reasonably satisfactory.'' 

Thirdly, one can capture between-study variation by running multi-site trials and conducting random effect meta-analysis or meta-regression. Some fields show a trend toward running multi-site trials~\citep{manylabs1,manylabs2}. 
Using unit-level data for such multi-site studies may allow a more fine-grained understanding of the discrepancies between studies.

\subsubsection*{Acknowledgments}

We thank Homa Zarghamee for sharing with us the student mood data reported in \cite{ifcher2011happiness}, Jonathan Taylor for suggesting the maximum selective likelihood approach, and Tian Zheng for helpful discussion on related literature and sharing with us the social network replication study data in~\cite{smith2023prediction}.

\bibliographystyle{plainnat}
\bibliography{bibliography.bib}

\newpage 
\appendix



\section{Properties of balancing estimators} \label{appendix:informal_properties}

This Appendix gives some high-level theoretical guidance on how the moments $\phi(x)$ and $\psi(m)$ should be selected in \Cref{section:estimators}'s estimators.  Since it is not our goal to derive new theory for ``balancing weights" estimators, we only present informal results without rigorous technical assumptions or proofs.  We refer the reader to the extensive balancing literature reviewed in \cite{benmichael2021balancing} for explicit technical conditions and worked-out derivations.

Our first informal proposition concerns the estimator $\widehat{\textsf{Covariate shift}}$ in the absence of publication bias.  It states that this estimator has an approximately normal sampling distribution centered at the true component \textsf{Covariate shift} under two possible assumptions.  The first of these assumptions concerns the $Q$-conditional average treatment effect (\ref{eq:Qcate}), and asks that it be a linear function of the balanced ``moments" $\phi(x)$.
\begin{align}
\tau_{\ell}(x) = \E_Q( Y_\ell \mid X = x, T = 1) - \E_Q ( Y_\ell \mid X = x, T = 0 ) \label{eq:Qcate}
\end{align}
The second assumption concerns the covariate densities $p_X(\cdot)$ and $q_X(\cdot)$, and asks that their log density ratio be linear in $\phi(x)$.

\begin{proposition}[Properties of $\widehat{\textsf{Covariate shift}}$] \label{proposition:covariate_shift_properties}
Assume that the observations in $\D_1$ and $\D_2$ are independent samples from $P$ and $Q$, respectively, and that the feature vector $\phi$ contains all functions of the form $x \mapsto \sum_{\ell = 1}^p f_{\ell,i}(x) \times f_{\ell,j}(x)$ as $i, j$ range over $\{ 1, \dots, \textup{dim}(f) \}$.  Assume that at least one of the following conditions holds:
\begin{enumerate}[label=(C\arabic*), itemsep=-0.5ex]
    \item $\sum_{\ell = 1}^p f_\ell(x) \tau_\ell(x) = \bm{B}_0 \phi(x)$ for some matrix $\bm{B}_0 \in \R^{\textup{dim}(\phi) \times p}$ and all $x \in \textup{support}(P_X)$. \label{item:linear_cate}
    \item $p_X(x) = \exp \{ \gamma_0^{\top} \phi(x) \} q_X(x)$ for some vector $\gamma_0 \in \R^{\textup{dim}(\phi)}$ and all $x \in \textup{support}(P_X)$. \label{item:loglinear_densityratio}
\end{enumerate}
Then, under suitable moment conditions, the approximation $\widehat{\textup{\textsf{Covariate shift}}} \, \dot{\sim} \, N( \textup{\textsf{Covariate shift}}, \frac{1}{n_1} \sigma_1^2 + \frac{1}{n_2} \sigma_2^2)$ holds for some $\sigma_1^2, \sigma_2^2$.
\end{proposition}

We explain the condition \ref{item:linear_cate} and \ref{item:loglinear_densityratio} by way of examples.  In \Cref{section:covid_misinfo}'s COVID news-sharing example, we had $p = 30$ and $f_{\ell}(x) = g_{\ell}(x) = \mathbf{1} \{ \text{Headline $\ell$ is true news} \}$.  Hence, \ref{item:linear_cate} would hold as long as the effect of treatment on ``willingness to share" \emph{all} headlines, $\sum_{\ell = 1}^{30} \tau_{\ell}(x)$, as well as the effect of treatment on ``willingness to share" \emph{true} headlines, $\sum_{\text{true headlines $\ell$}} \tau_{\ell}(x)$, both vary linearly with the balanced features.  Meanwhile, \ref{item:loglinear_densityratio} holds if, say, the covariate $X$ has a normal distribution in both $P$ and $Q$ and we take $\phi(x) = (1, x, x^2)$.  This was the case in \Cref{section:stylized_examples}'s stylized examples.  

While neither condition is likely to hold exactly, we will mention that the normal approximation also remains valid when both conditions hold approximately.  Even if neither condition holds approximately, the sampling distribution of $\widehat{\textsf{Covariate shift}}$ remains approximately normal, although centered around a different quantity:  the amount by which the population discrepancy would decline had the replication experiment instead been conducted in the ``nearest" distribution to $Q$ which matches $P$ in terms of the moments $\phi(X)$.  

Our next proposition provides similar conditions under which $\widehat{\textsf{Mediation shift}}$ has an approximately normal sampling distribution centered at the true value of \textsf{Mediation shift}.  Since $\textsf{Mediation shift}$ is a more complex estimand than $\textsf{Covariate shift}$, the required assumptions are correspondingly stronger.  Here, we must impose a linearity assumption not only on the treatment effect $\tau_{\ell}$, but also the entire outcome regression:
\begin{align}
\mu_{\ell}(x, m, t) &= \E_Q(Y_{\ell} \mid X = x, M = m, T = t)
\end{align}

\begin{proposition}[Properties of $\widehat{\textsf{Mediation shift}}$] \label{prop:mediation_shift}
Assume the conditions of \Cref{proposition:covariate_shift_properties} (including either \ref{item:linear_cate} or \ref{item:loglinear_densityratio}).  Further suppose that one of the following conditions holds for both $t = 0$ and $t = 1$:
\begin{enumerate}[itemsep=-0.5ex, label=(M\arabic*)]
    \item $\sum_{\ell = 1}^p f_{\ell}(x) \mu_{\ell}(x, m, t) = \bm{U}_t \phi(x) + \bm{V}_t \psi(m)$ for some matrices $\bm{U}_t \in \R^{\textup{dim}(\phi) \times p}, \bm{V}_t \in \R^{\textup{dim}(\psi) \times p}$. \label{item:linear_outcome_model_mediator}
    \item $p_{M \mid X, T}(m \mid x, t) = \exp \{ \eta_t^{\top} \psi(m) \} q_{M \mid X, T}(m \mid x, t)$ and \ref{item:loglinear_densityratio} holds. 
    \label{item:loglinear_densityratio_mediator}
\end{enumerate}
Then, under suitable moment conditions, the approximation $\widehat{\textsf{\textup{Mediation shift}}} \, \dot{\sim} \, N(\textsf{\textup{Mediation shift}}, \frac{1}{n_1} \tau_1^2 + \frac{1}{n_2} \tau_2^2)$ holds for some $\tau_1^2, \tau_2^2$.
\end{proposition}

\section{Details of selection-adjusted point estimators} \label{appendix:selective_mle}

In this section, we motivate and describe the selection-adjusted point estimators mentioned in \Cref{section:selection_adjustment} of the main text.  These estimators are obtained by approximately maximizing the conditional-on-replication distribution of \Cref{section:estimators}'s estimators.

In the absence of selection bias, \Cref{appendix:informal_properties} describes conditions under which our component estimators have approximately normal distributions centered at the true components:
\begin{align}
\left[ 
\begin{array}{c}
\hat{\tau}/\hat{\sigma}\\
\theta(\D_1) - \theta(\D_2)\\
\widehat{\textsf{Covariate shift}}\\
\widehat{\textsf{Mediation shift}}
\end{array}
\right] \quad \dot{\sim} \quad N \left( \left[ 
\begin{array}{c}
\eta\\
\theta(P) - \theta(Q)\\
\textsf{Covariate shift}\\
\textsf{Mediation shift}
\end{array}
\right], 
\bm{\Sigma}
\right) \label{eq:approximate_joint_normality}
\end{align}
Here, $\hat{\tau}$ is the regression coefficient whose significance level governs the chance that the original study is published/replicated, and $\hat{\sigma}$ is its standard error.  Meanwhile, $\eta$ is the true standardized effect size in the original study.

Under selection bias of the type imposed in \Cref{assumption:selection}, the conditional-on-replication distribution of our estimates is no longer approximately normal.  Instead, the following (informal) Lemma says that the distribution will be approximately \emph{truncated} normal.

\begin{lemma}[Conditional-on-replication distribution is truncated normal] \label{lemma:truncated_normal}
Let $\hat{\Delta}$ denote the vector on the left-hand side of \Cref{eq:approximate_joint_normality}, and let $\Delta$ denote the mean of the normal distribution on the right-hand side.  If $\hat{\Delta} \, \dot{\sim} \, N(\Delta, \bm{\Sigma})$ in the absence of selection bias by journals or replicators, and selection bias operates in the manner prescribed by Assumptions \ref{assumption:selection} and \ref{assumption:replication_sample_size}, then the following approximation holds for any set $A$:
\begin{align}
\P( \hat{\Delta} \in A \mid \textup{Original study published/replicated with } {p} < \alpha_0) \approx \P_{Z \sim N(\Delta, \bm{\Sigma})}(Z \in A \mid |Z_1| > z_{1 - \alpha_0/2}). \label{eq:truncated_gaussian_approximation}
\end{align}
\end{lemma}

Based on this lemma, we could estimate the vector $\Delta$ by maximizing the likelihood of the truncated Gaussian distribution on the right-hand side of \Cref{eq:truncated_gaussian_approximation}. 

However, we can slightly improve over this approach by exploiting the fact that $\Delta_1 = \eta$ is irrelevant for our decomposition.  We therefore eliminate this nuisance parameter from our likelihood by defining $R = \hat{\Delta}_1 - \bm{\Sigma}_{1, 2:4}^{\top} \bm{\Sigma}_{2:4, 2:4}^{-1} \hat{\Delta}_{2:4}$ and noticing that the (approximate) conditional distribution of $\hat{\Delta}_{2:4}$ given $R$ has a density not depending on $\Delta_1$:
\begin{align*}
\textup{Likelihood}(\Delta; \hat{\Delta}_{2:4}) &:= \P_{Z \sim N(\Delta, \bm{\Sigma})}( Z_{2:4} = \hat{\Delta}_{2:4} \mid R = \hat{\Delta}_1 - \bm{\Sigma}_{1, 2:4} \bm{\Sigma}_{2:4, 2:4}^{-1} \hat{\Delta}_{2:4}, |Z_1| > z_{1 - \alpha_0/2})\\
&= \P_{Z \sim N(\Delta, \bm{\Sigma})}(Z_{2:4} = \hat{\Delta}_{2:4} \mid | \hat{\Delta}_1 + \bm{\Sigma}_{1, 2:4}^{\top} \bm{\Sigma}_{2:4, 2:4}^{-1} (Z_{2:4} - \hat{\Delta}_{2:4})| > z_{1 - \alpha_0/2})\\
&\propto \frac{\exp \{ -\frac{1}{2} (\hat{\Delta}_{2:4} - \Delta_{2:4})^{\top} \bm{\Sigma}_{2:4, 2:4}^{-1} (\hat{\Delta}_{2:4} - \Delta_{2:4}) \}}{\P_{Z_{2:4} \sim N(\Delta_{2:4}, \bm{\Sigma}_{2:4, 2:4})}( |\hat{\Delta}_1 + \bm{\Sigma}_{1,2:4}^{\top} \bm{\Sigma}_{2:4, 2:4}^{-1} (Z_{2:4} - \hat{\Delta}_{2:4})| > z_{1 - \alpha_0/2})}.
\end{align*}

Our selection-adjusted point estimators for $\textsf{Covariate shift}$ and $\textsf{Mediation shift}$, as well as the population discrepancy $\theta(P) - \theta(Q)$, are obtained by maximizing the (logarithm of) the final-line in the preceding display with the unknown covariance $\bm{\Sigma}$ replaced by a jackknife estimate.  This is an unconstrained three-dimensional optimization problem, which we solve using the Nelder-Mead algorithm as implemented by the \texttt{optim()} function in \textsf{R}.

Finally, we estimate $\textsf{Sampling variability}$ as the difference between the observed discrepancy $\theta(\D_1) - \theta(\D_2)$ and our selection-adjusted estimate of $\theta(P) - \theta(Q)$.  We obtain a selection-adjusted estimate of \textsf{Residual factors} by enforcing the (deterministic) relationship $\theta(P) - \theta(Q) = \textsf{Covariate shift} + \textsf{Mediation shift} + \textsf{Residual factors}$ on our point estimators.

\section{Simulations for power-calculated sample size} \label{appendix:simulations}

We conduct simulations to illustrate the validity of our methods when the replication sample size $n_2$ is obtained via a power calculation based on the original study's point estimate, thereby violating \Cref{assumption:replication_sample_size}. As mentioned in the main text, we are to show the limited impact of such deviation from independent sampling. As a baseline, we also evaluate our methods for standard independently sampled original and replication datasets in order to verify our theory. 

\subsection{Simulation setups}

\paragraph{Data generating processes.}

We design settings that vary in how $(X,M,Y)$ are related and thus the takeaway message for the role of distribution shifts.  Within each generic setting, we design sub-settings  that obey three types of identification conditions (c.f.~Appendix): (i) both density and outcome relationship are (log-)linear, (ii) only outcome relationship is linear, and (iii) only density is log-linear. 
The double robustness result in Appendix imply that our method is valid in all the three cases.  

Throughout this section, we consider data generating processes such that
\begin{align}
\label{eq:simu_general}
Y = \mu_0(X,M) + T\cdot \delta(X,M) + \sigma \cdot \cN(0,1),
\end{align}
where $\sigma>0$ is the noise scale, 
$\mu_0(x,m)$ is the baseline mean function (i.e., $\EE[Y \given T = 0, X=x,M=m]$), 
and $\delta(x,m)$ is the conditional average treatment effect (CATE) function. 
For clarity, 
we denote the $\mu_0$ and $\delta$ functions under the original distribution $P$ and the replication distribution $Q$ by $\mu_0(x,m;P)$, $\mu_0(x,m;Q)$, $\delta(x,m;P)$ and $\delta(x,m;Q)$. 
We introduce detailed   distribution of $(X,M)$ and  specification of $\mu_0$, $\delta$, and $\sigma$ in later subsections.

\paragraph{Evaluated methods.} 

For each data generating process, we evaluate the following procedures:
\begin{itemize}
    \item \emph{Standard approach}. Independently draw $n_1$ samples from $P$ and $n_2$ samples from $Q$ as the original study and replication study, respectively, and conduct our decomposition estimation with entropy balancing. 
    \item \emph{Power-calculated approach}. We first independently draw $n_1$ samples from $P$ as the data for the original study and compute the effect $\theta(P_{n_1})$, then calculate $n_2$ as the smallest integer such that the power for detecting $90\%$ of the estimated effect size $\theta(P_{n_1})$ is $0.9$; this calculation uses $\theta(P_{n_1})$ and the estimated standard deviation of $\theta(P_{n_1})$. We then generate $n_2$ i.i.d.~sample from $Q$ as the data for the replication study, and conduct our method as if the datasets were sampled independently.
\end{itemize} 
To explore the performance of the power-calculated approach under different power regimes, we also vary the signal-to-noise ratio in each data generating process. 
The procedures are repeated independently for $500$ times, and the empirical coverage for $90\%$ confidence intervals are evaluated. 

\subsection{Setting 1: Pure observed distribution shift}
\label{app:simu_set1}

We first consider settings where the observed discrepancy is entirely attributed to observed distribution shift. 
To be specific, we let $\delta(x,m;P)=\delta(x,m;Q)=\delta(x,m)$ for some fixed function $\delta(x,m)$, and $\mu_0(x,m;P)=\mu_0(x,m;Q)\equiv 0$. In this way, the ground truth of the residual factor is always zero. 
The entropy balancing features are always $\phi(x)=x$ for the covariate shift and $\psi(x) = (x,m)$ for the mediation shift. 

In this setting, we enforce $(X,M)$ to be mutually independent.
We generate settings with three types of identification conditions\footnote{Throughout, we use $\cN(\cdot,\cdot)$ to denote (mutually independently generated) Gaussian random variables.}: 
\begin{enumerate} 
    \item[(i)] Both linear: $\delta(x,m)=x+m$. Under $P$, we set $X\sim \cN(1/2,1)$ and $M\sim \cN(1/2,1)$. Under $Q$, $X\sim \cN(0,1)$ and $M\sim \cN(0,1)$.
    \item[(ii)] Only outcome is linear: $\delta(x,m)=x+m$. Under $P$, we set $X\sim 1/2\cdot \cN(1,1)+1/2\cdot \cN(0,1)$ and $M\sim \cN(1/2,1)$. Under $Q$, $X\sim \cN(0,1)$ and $M\sim \cN(0,1)$. In this way, the density ratio $\ud P_X/\ud Q_X$ is log-linear in $(x,x^2)$, but we only use $\phi(x)=x$ for entropy balancing.  
    \item[(iii)] Only density is log-linear: $\delta(x,m) = 1.1(x+m+x^2/2+m^2/2)$. Under $P$, we set $X\sim \cN(1/2,1)$ and $M\sim \cN(1/2,1)$. Under $Q$, $X\sim \cN(0,1)$ and $M\sim \cN(0,1)$. The CATE function for the covariate is linear in $(x,x^2)$, which is misspecified for $\phi(x)=x$. The function $\delta$ is also misspecified for $\psi(m)=m$.
\end{enumerate}

For the standard approach, we fix the sample sizes $n=N=500$. For the power-calculated approach, we fix the original study sample size at $n=500$ and vary the noise level $\sigma\in \{1, 1.5, 2, \dots, 3\}$, which leads to different  distributions of $N$. 

\begin{figure}[htbp]
    \centering
    \includegraphics[width=5.5in]{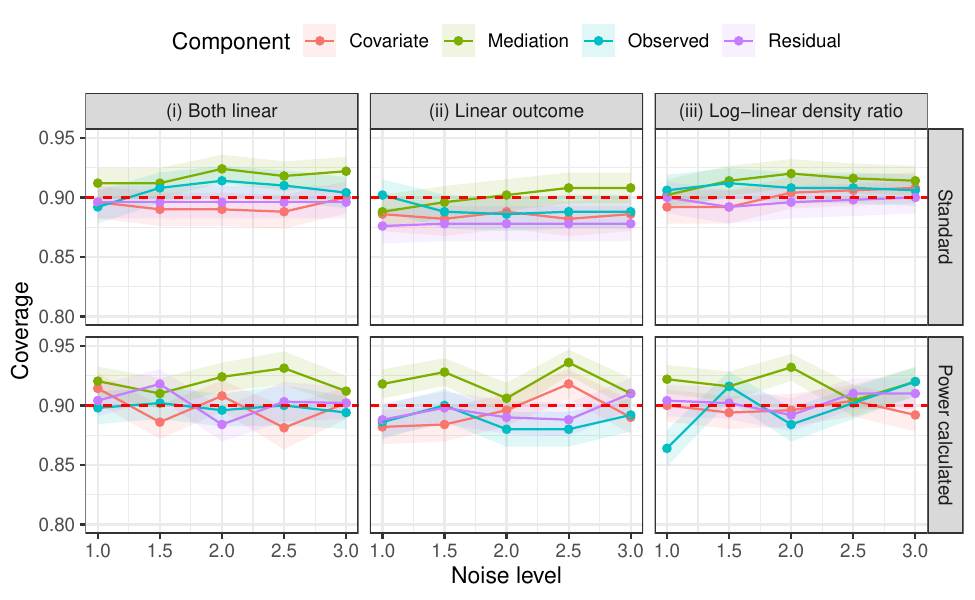}
    \caption{Empirical coverage ($\pm$ standard deviation) of confidence intervals with nominal coverage level $90\%$ for three components (Covariate, Mediator, Residual) and the total discrepancy (Observed). Results are evaluated over $500$ independent repeats in three sub-settings in Section~\ref{app:simu_set1}. The $x$-axis is the noise level $\sigma$ in~\eqref{eq:simu_general}.}
    \label{fig:simu_simple}
\end{figure}

The empirical coverage of $90\%$ confidence intervals 
for three components and the total discrepancy is summarized in Figure~\ref{fig:simu_simple}
for the two evaluated approaches. 
The first row of Figure~\ref{fig:simu_simple} verifies the validity of our confidence intervals and its double robustness under various noise levels. We observe slightly increasing coverage as $\sigma$ increases. 
The second row of Figure~\ref{fig:simu_simple} justifies our claim that the power calculation, regardless of the signal strength hence the relative scale of $n_2$ (the average sample size $n_2$ under various noise levels is summarized in Table~\ref{tab:set1_N}), does not impair the validity of our confidence intervals. 

\begin{table}[ht]
\centering
\renewcommand{\arraystretch}{1.2}
\begin{tabular}{c|ccccc}
  \toprule
Setting & $\sigma=1$ & $\sigma=1.5$ & $\sigma=2$ & $\sigma=2.5$ & $\sigma=3$  \\ 
  \hline 
  (i) Both linear 
  & 119.6 & 182.2& 294.5& 454.9  & 690.4 \\ 
    \hline
 (ii) Linear outcome & 126.4 & 192.4 & 314.3 & 498.0 & 752.4 \\ 
    \hline
(ii) Log-linear density & 109.6 & 219.9 & 401.3& 646.2& 1040.3 \\
   \bottomrule
\end{tabular}
\caption{Sample mean of $n_2$ for the power-calculated approach. Results are evaluated over $500$ independent repeats in three sub-settings in Section~\ref{app:simu_set1}. $\sigma$ is the noise level in~\eqref{eq:simu_general}.}
\label{tab:set1_N}
\end{table}

\subsection{Setting 2: Pure observed distribution shift, multivariate}
\label{app:simu_set2}

We then design a more complicated setting where $(X,M)$ could be mutually dependent and multivariate. We will allow for some misspecification for these relatively complex settings to check the robustness of our approach. Still, $\delta(x,m;P)=\delta(x,m;Q)=\delta(x,m)$ for a fixed function $\delta$. We take $X\in \RR^4$ and $M\in \RR^3$, and generate three types of identification conditions:

\begin{enumerate} 
    \item[(i)] Both linear: $\delta(x,m)=x_1+x_3+m_1$. Under $P$, $X\sim \cN(\mu_x,I_4)$ for $\mu_x=(1/2,-1/2,0,0)^\top$ and $M$ is generated via $M_1 = \cN(0,1/4) +  X_1/2 + 1/2$, $M_2 = \cN(0,1/4) +  X_1/2 - 1/2$, $M_3 = \cN(0,1)$. Under $Q$, we set $X\sim \cN(0,I_4)$ and $M\sim \cN(0,I_3)$. 
We use $\phi(x) = (x_1,x_2,x_3)$, $\psi(m)=(m_1,m_2)$ for entropy balancing. In this case, the outcome relation is linear for both covariates and mediators, but the density ratio is log-linear only for the covariates (so there is slight misspecification).
    \item[(ii)] Only outcome is linear: $\delta(x,m)=x_1+x_3+m_1$. Under $P$, we first generate $I^0\sim \textrm{Bern}(1/2)$, $X^0\sim \cN(0,I_4)$, and set $X_1=X_1^0+I^0/2$, $X_2=X_2^0-I^0/2$, and $X_{3:4}=X_{3:4}^0$. Then, we generate $M_0\sim \cN(0,I_3)$, and take $M_1=M_1^0+T$, $M_2=M_2+0.5X_1-0.5$ to create dependency. Under $Q$, $X\sim \cN(0,I_4)$ and $M\sim \cN(0,I_3)$. We use $\phi(x) = (x_1,x_2,x_3)$, $\psi(m)=(m_1,m_2)$ for entropy balancing. In this way, both $\phi(x)$ and $\psi(m)$ are misspecified for the density ratio. 
    \item[(iii)] Only density is log-linear:  $\delta(x,m) = x_1 + x_3^2/4 + m_1^2$. 
    Under $P$, $X\sim \cN(\mu_x, I_4)$ for $\mu_x = (1/2,-1/2, 0,0)^\top$, and $M$ is generated via $M_1 = \cN(0,1/4)+ X_1/2$, $M_2 = \cN(0, 1/4) + X_1/2-1/2$, and $M_3\sim \cN(0,1)$. 
    Under $Q$, $X\sim \cN(0,I_4)$ and $M\sim \cN(0,I_3)$. For entropy balancing, the covariate shift component uses  $\phi(x) = (x_1,x_3)$, while the mediation part uses $\psi(m)=(m_1,m_2,m_1^2,m_2^2)$. In this case, the outcome model is misspecified. The density ratio is also slightly misspecified: it is log-linear also in $(x_1m_1,x_1m_2)$ which is left out from our features. 
\end{enumerate}

\begin{figure}[htbp]
    \centering
    \includegraphics[width=5.5in]{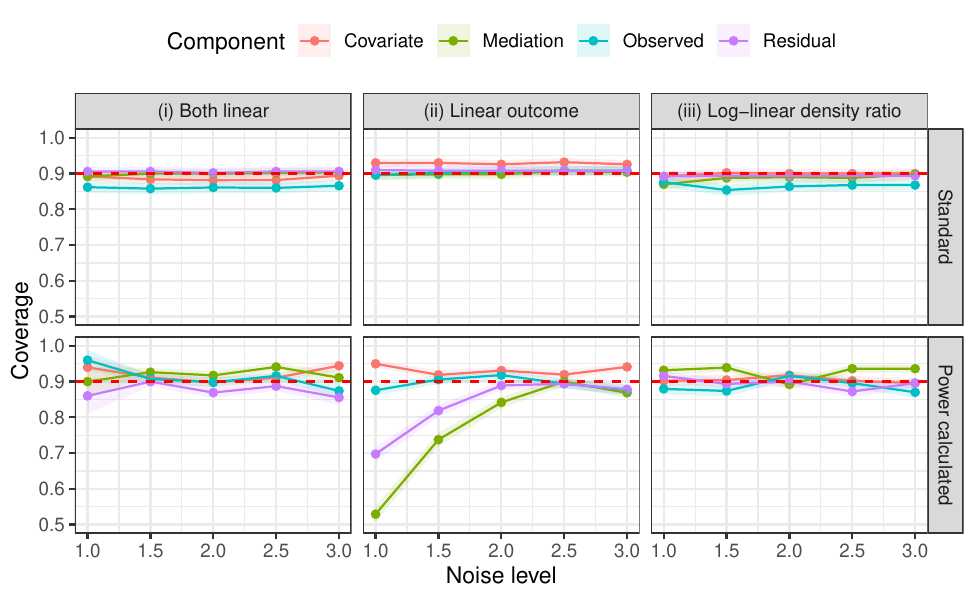}
    \caption{Empirical coverage ($\pm$ standard deviation) of confidence intervals with nominal coverage level $90\%$ for three components (Covariate, Mediator, Residual) and the total discrepancy (Observed). Results are evaluated over $500$ independent repeats in three sub-settings in Section~\ref{app:simu_set2}. The $x$-axis is the noise level $\sigma$ in~\eqref{eq:simu_general}.}
    \label{fig:simu_multi}
\end{figure}

The empirical coverage of $90\%$ confidence intervals for all components and observed discrepancy is summarized in Figure~\ref{fig:simu_multi}, with the corresponding average $n_2$ for the power-calculated approach in Table~\ref{tab:set2_N}.

With fixed sample sizes $n_1=n_2=500$, the standard approach achieves the nominal coverage for all components, at all noise levels, in all settings. Notably, this is despite the fact that the identification conditions are sometimes violated: recall that in setting (iii), even the density ratio has slight misspecification. This means our method is robust even when the identification conditions are not perfectly fulfilled. 

On the other hand, for small noise level ($\sigma<2$), the power-calculated approach undercovers in setting (ii), which corresponds to an average $n_2 < 150$ in Table~\ref{tab:set2_N}. 
However, when the sample size is reasonably large (larger than half of $n_1$), the coverage becomes good again. As the replication study usually has a larger sample size than the original study, we expect the power calculation to have little impact on the validity of our approach in most cases in practice. 

\begin{table}[ht]
\centering
\renewcommand{\arraystretch}{1.2}
\begin{tabular}{c|ccccc}
  \toprule
Setting & $\sigma=1$ & $\sigma=1.5$ & $\sigma=2$ & $\sigma=2.5$ & $\sigma=3$  \\ 
  \hline 
  (i) Both linear 
 & 99.62 & 144.28 & 215.62 & 318.04 & 459.01 \\ 
    \hline
 (ii) Linear outcome & 88.23 & 136.14 & 204.60 & 302.97 & 444.19 \\ 
    \hline
(iii) Log-linear density & 65.89 & 107.54 & 169.56 & 255.98 & 378.10 \\ 
   \bottomrule
\end{tabular}
\caption{Sample mean of $n_2$ for the power-calculated approach. Results are evaluated over $500$ independent repeats in three sub-settings in Section~\ref{app:simu_set2}. $\sigma$ is the noise level in~\eqref{eq:simu_general}.}
\label{tab:set2_N}
\end{table}

\subsection{Setting 3: Residual distribution shift}
\label{app:simu_set3}
We finally design settings with nontrivial residual shift, i.e., the conditional distributions of the outcomes given $(X,M,T)$ differ across studies. 
We still follow~\eqref{eq:simu_general} to generate the data, where we use $\delta(x,m;P)\neq \delta(x,m;Q)$ and $\mu_0(x,m;P)=\mu_0(x,m;Q)\equiv 0$. 

We design settings that allow for nontrivial residual factors. 
In specific, we take $\delta(x,m;Q)$ to be the same as the choices in Section~\ref{app:simu_set2}, while setting a distinct $\delta(x,m;P)$. 
This injects slight model misspecification to test the robustness of our methods. 
We take $X\in \RR^4$ and $M\in \RR^3$, and generate three types of identification conditions, all with slight misspecification:

\begin{enumerate} 
    \item[(i)] Both linear: $\delta(x,m;Q)=x_1+x_3+m_1$, $\delta(x,m;P)=x_1+x_3+x_1^2/2$. Under $P$, $X\sim \cN(\mu_x,I_4)$ for $\mu_x=(1/2,-1/2,0,0)^\top$ and $M$ is generated via $M_1 = \cN(0,1/4) +  X_1/2 + 1/2$, $M_2 = \cN(0,1/4) +  X_1/2 - 1/2$, $M_3 = \cN(0,1)$. Under $Q$, we set $X\sim \cN(0,I_4)$ and $M\sim \cN(0,I_3)$. 
We use $\phi(x) = (x_1,x_2,x_3)$, $\psi(m)=(m_1,m_2)$ for entropy balancing. 
    \item[(ii)] Only outcome is linear: $\delta(x,m;Q)=x_1+x_3+m_1$, $\delta(x,m;P)=x_1+x_3+0.7$. Under $P$, we first generate $I^0\sim \textrm{Bern}(1/2)$, $X^0\sim \cN(0,I_4)$, and set $X_1=X_1^0+I^0/2$, $X_2=X_2^0-I^0/2$, and $X_{3:4}=X_{3:4}^0$. Then, we generate $M_0\sim \cN(0,I_3)$, and take $M_1=M_1^0+T$, $M_2=M_2+0.5X_1-0.5$. Under $Q$, $X\sim \cN(0,I_4)$ and $M\sim \cN(0,I_3)$. We use $\phi(x) = (x_1,x_2,x_3)$, $\psi(m)=(m_1,m_2)$.
    \item[(iii)] Only density is log-linear:  $\delta(x,m;Q) = x_1 + x_3^2/4 + m_1^2$, $\delta(x,m;P) = 2x_1$. 
    Under $P$, $X\sim \cN(\mu_x, I_4)$ for $\mu_x = (1/2,-1/2, 0,0)^\top$, and $M$ is generated via $M_1 = \cN(0,1/4)+ X_1/2$, $M_2 = \cN(0, 1/4) + X_1/2-1/2$, and $M_3\sim \cN(0,1)$. 
    Under $Q$, $X\sim \cN(0,I_4)$ and $M\sim \cN(0,I_3)$. For entropy balancing, the covariate shift component uses  $\phi(x) = (x_1,x_3)$, while the mediation part uses $\psi(m)=(m_1,m_2,m_1^2,m_2^2)$. 
\end{enumerate}

\begin{figure}[htbp]
    \centering
    \includegraphics[width=5.5in]{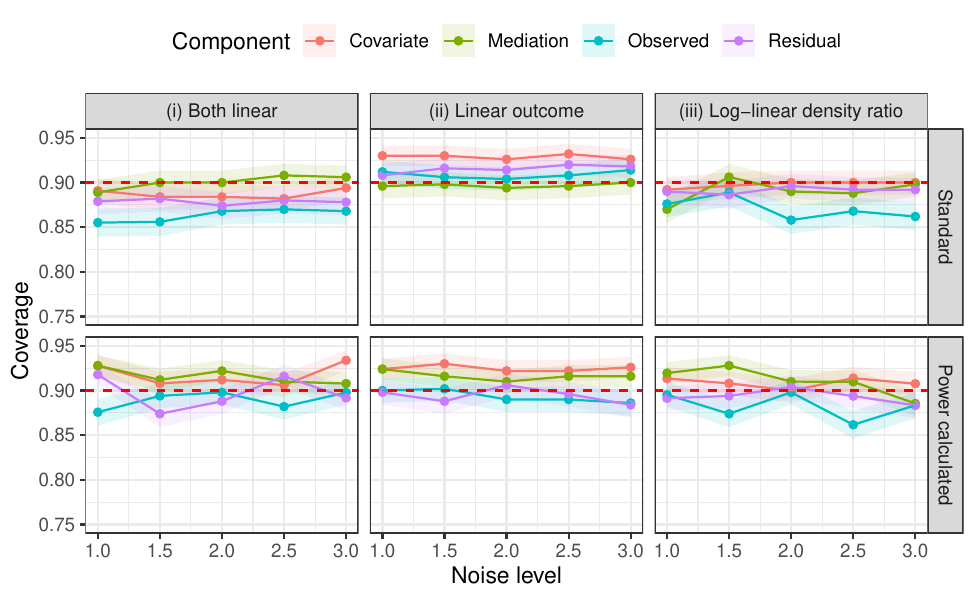}
    \caption{Empirical coverage ($\pm$ standard deviation) of confidence intervals with nominal coverage level $90\%$. Results are evaluated over $500$ independent repeats in three sub-settings in Section~\ref{app:simu_set3}. The $x$-axis is the noise level $\sigma$ in~\eqref{eq:simu_general}.}
    \label{fig:simu_resid}
\end{figure}

The empirical coverage of $90\%$ confidence intervals is summarized in Figure~\ref{fig:simu_resid}. It confirms the validity of the standard approach and the power-calculated approach across all noise levels and identification conditions. 
The average sample size $n_2$ is reported in Table~\ref{tab:set3_N}.

\begin{table}[ht]
\centering
\renewcommand{\arraystretch}{1.2}
\begin{tabular}{c|ccccc}
  \toprule
Setting & $\sigma=1$ & $\sigma=1.5$ & $\sigma=2$ & $\sigma=2.5$ & $\sigma=3$  \\ 
  \hline 
   (i) Both linear  & 124.03 & 184.96 & 278.84 & 419.15 & 618.10 \\ 
  (ii) Linear outcome  & 139.92 & 208.29 & 341.79 & 557.40 & 838.81 \\ 
  (iii) Log-linear density & 167.17 & 249.42 & 380.48 & 564.57 & 862.41 \\  
   \bottomrule
\end{tabular}
\caption{Sample mean of $n_2$ for the power-calculated approach for settings in Section~\ref{app:simu_set3}.}
\label{tab:set3_N}
\end{table}

\section{Simulations for selection-adjusted inference}

We then conduct simulations to demonstrate the validity of our post-selection inference that accounts for publication bias. We continue to use the settings in the preceding section, but only keep the original study if its p-value for the $T$ coefficient is smaller than $0.05$. 

Specifically, we adopt the three data generating processes in Appendix~\ref{app:simu_set2}; they are more challenging than those in Appendix~\ref{app:simu_set1} and similar to those in Appendix~\ref{app:simu_set3}. We slightly change the formulation to
\#\label{eq:dgp_selective}
Y = \mu_0(X,M) + \nu \cdot T\cdot \delta(X,M) + \cN(0,1),
\#
where $\nu\in\{0.05, 0.1, 0.15,0.2\}$ is the signal strength, and the choice of $\mu_0$ and $\delta$ is the same as Appendix~\ref{app:simu_set2}. 

Fixing $n_1=500$, we repeatedly sample $n_1$ i.i.d.~data $(X_i,M_i,Y_i,T_i)_{i=1}^{n_1}$ from $P$ until the p-value for the $T$ coefficient in the regression $Y\sim T$ is below $0.05$. Then, we calculate $n_2$ as the smallest integer such that the power for detecting $90\%$ of the estimated effect size $\theta(P_{n_1})$ is $0.8$. We then sample $n_2$ i.i.d.~data from $Q$ as the replication study data. 

\begin{figure}[htbp]
    \centering
    \includegraphics[width=5.5in]{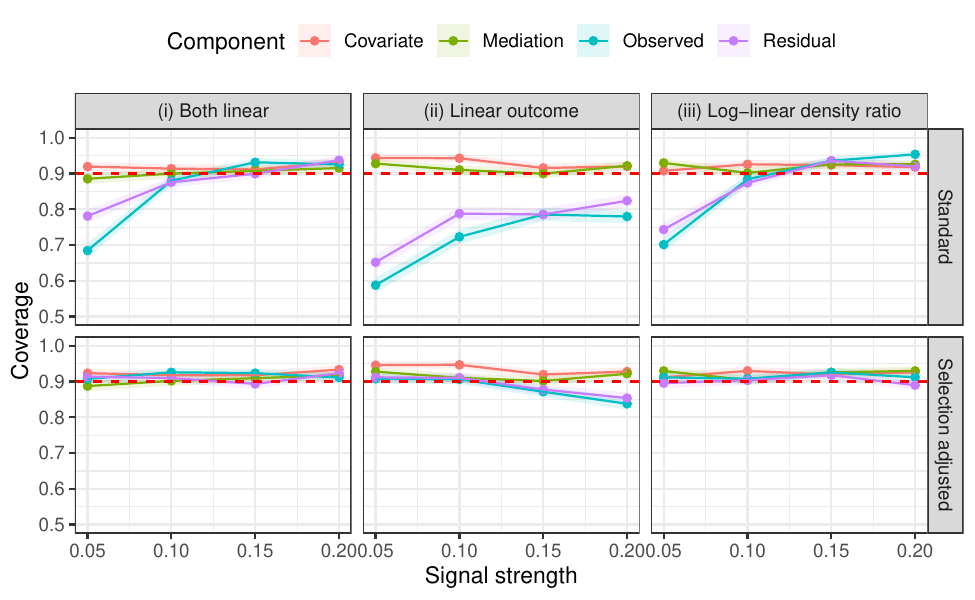}
    \caption{Empirical coverage ($\pm$ standard deviation) of confidence intervals with nominal coverage level $90\%$ for three components (Covariate, Mediator, Residual) and the total discrepancy (Observed) when publication bias occurs for $p<0.05$. The first row plots the coverage for the standard approach, and the second row is for the selection adjustment.  Results are evaluated over $500$ independent repeats. The $x$-axis is the signal strength $\nu$ in~\eqref{eq:dgp_selective}.}
    \label{fig:simu_selective}
\end{figure}

Using the two datasets, we evaluate two methods: 
our standard approach in Section~\ref{section:estimators} and the selection-adjusted approach in Section~\ref{section:selection_adjustment}; we 
compute the empirical coverage of confidence intervals for the three components and the observed discrepancy. 
The results across three settings are summarized in Figure~\ref{fig:simu_selective}. 

In Figure~\ref{fig:simu_selective}, the standard approach suffers from selection bias in all settings, especially for the residual component and the observed discrepancy. In contrast, our selection adjusted confidence intervals maintain valid and tight coverage under all values of $\alpha$ and in all settings. 
Interestingly, the covariate shift and mediation shift components are less prone to selection bias on the $T$ coefficient; we conjecture that this is because in these settings, the influence function for the two estimators are approximately independent of the influence function for the $T$ coefficient in the original regression. 

\section{Proof sketches}

This section provides proof sketches for the informal theoretical properties stated in \Cref{appendix:informal_properties}. 
The proof ideas are based on well-establish theory for entropy balancing. While we still need slight extension of existing proof techniques to handle linear regression with multivariate outcomes or interactions, we omit most technical details for brevity. 

Throughout these proof sketches, we use the notation $\hat{\E}_{P \times Q}$ to denote the two-sample U-statistic operator:
\begin{align*}
\hat{\E}_{P \times Q}[ f(X_i, Y_i, M_i, T_i, X_j, Y_j, M_j, T_j)] &= \frac{1}{n_1 n_2} \sum_{\substack{i \in \D_1\\j \in \D_2}} f(X_i, Y_i, M_i, T_i, X_j,  Y_j, M_j, T_j)
\end{align*}

\subsection{Proof sketch for \Cref{proposition:covariate_shift_properties}}
\begin{proof}[Proof sketch for \Cref{proposition:covariate_shift_properties}]
Let $(\theta_w(\cD_2), \beta_w(\cD_2))$ denote the WLS coefficients computed on $\cD_2$ with the entropy-balancing weights $\hat{w}$, that is, $\theta_w(\cD_2)=\theta(\D_2, \textup{weights} = \hat{w})$ defined in our main text, and similarly for $\beta_w(\cD_2)$.  By convex duality, the weights $\hat{w}$ will take the form $\hat{w}_i = n_2^{-1} \exp \{ \hat{\gamma}_0^{\top} \phi(X_i) + T_i \hat{\gamma}_1^{\top} \phi(X_i) \}$ for some vector $\hat{\gamma} = (\hat{\gamma}_0, \hat{\gamma}_1) \in \R^{2 \, \textup{dim}(\phi)}$.  Therefore, the estimator $\widehat{\textsf{Covariate shift}}$ can be characterized as the solution to a U-statistic estimating equation:
\begin{align*}
\hat{\E}_{P \times Q}[ u( X_i, Y_i, T_i, M_i, X_j, Y_j, T_j, M_j, \theta(\D_1), \beta(\D_1), \theta(\D_2), \beta(\D_2), \hat{\gamma}, \theta_w(\D_2), \beta_w(\D_2), \widehat{\textsf{Covariate shift}})] = 0
\end{align*}
where the moment function $u(\cdot)$ is given by:
\begin{align}
\begin{split}
&u(x_i, y_i, t_i, m_i, x_j, y_j, t_j, m_j, \theta_1, \beta_1, \theta_2, \beta_2, \gamma, \theta_w, \beta_w, \Delta)\\
&\quad = \left[ 
\begin{array}{c}
\sum_{\ell = 1}^p [y_{i\ell} - (\theta_1, \beta_1)^{\top}(\tilde t_i f_{\ell}(x_i), g_{\ell}(x_i))] (\tilde t_i f_{\ell}(x_i), g_{\ell}(x_i))\\
\sum_{\ell = 1}^p [y_{j \ell} - (\theta_2, \beta_2)^{\top} (\tilde t_j f_{\ell}(x_j), g_{\ell}(x_j))](\tilde t_j f_{\ell}(x_j), g_{\ell}(x_j))\\
\exp \{ \gamma^{\top} ( \phi(x_j), t_j \phi(x_j)) \} \phi(x_j) - \phi(x_i)\\
\exp \{ \gamma^{\top}( \phi(x_j), t_j \phi(x_j) \} t_j \phi(x_j) - t_i \phi(x_i)\\
\sum_{\ell = 1}^p \exp\{ \gamma^{\top} ( \phi(x_j), t_j \phi(x_j)) \} [ y_{j \ell} - (\theta_w, \beta_w)^{\top} (\tilde t_j f_{\ell}(x_j), g_{\ell}(x_j))] (\tilde t_j f_{\ell}(x_j), g_{\ell}(x_j))\\
(\theta_1 - \theta_2) - (\theta_1 - \theta_w) - \Delta
\end{array}
\right].
\end{split} \label{eq:covariate_moment_equation}
\end{align}
The interpretation of the moment equation $u$ is as follows.  The first equation is the OLS first-order condition identifying $(\theta(\D_1), \beta(\D_1))$, where we have centered the treatment by defining $\tilde t_i = t_i - P(T = 1)$.  Since $g_{\ell}$ contains $f_{\ell}$ as a subvector, this centering has no effect on the regression coefficient $\theta_1$.  The second equation in $u$ is the OLS first-order condition identifying $(\theta(\D_2), \beta(\D_2))$.  Once again, we have centered the treatment by setting $\tilde t_j = t_j - Q(T = 1)$.  The third and fourth equations are the balancing conditions identifying $\hat{\gamma}$ (and hence $\hat{w}$)~\citep{zhao2017entropy}, the fifth equation is the WLS first-order condition identifying $(\theta_w(\D_2), \beta_w(\D_2))$, and the last equation defines $\widehat{\textsf{Covariate shift}}$.

By U-statistic Z-estimator theory \citep{honore1994pairwise}, the sample Z-estimators will have an approximately normal sampling distribution centered at the solution to the \emph{population} moment equation:
\begin{align*}
\left[ 
\begin{array}{c}
\theta(\D_1)\\
\beta(\D_1)\\
\theta(\D_2)\\
\beta(\D_2)\\
\hat{\gamma}\\
\theta_w(\D_2)\\
\beta_w(\D_2)\\
\widehat{\textsf{Covariate shift}}
\end{array}
\right] \quad \dot{\sim} \quad N \left( 
\left[ 
\begin{array}{c}
\theta_1^*\\
\beta_1^*\\
\theta_2^*\\
\beta_2^*\\
\gamma^*\\
\theta_w^*\\
\beta_w^*\\
\Delta^*
\end{array}
\right], \frac{1}{n_1} \bm{\Sigma}_1 + \frac{1}{n_2} \bm{\Sigma}_2
\right).
\end{align*}
We will also mention that if we write $\gamma^*$ as $(\gamma_0^*, \gamma_1^*)$, then $\gamma_1^*$ must be zero, as the population entropy balancing solution which balances $\phi(X)$ will automatically balance $T \phi(X)$.

We now argue that $\Delta^* = \textsf{Covariate shift}$ holds under either of the two conditions stated in the Proposition.  As long as there is no feature collinearity under either $P$ or $Q$, the equalities $\theta_1^* = \theta(P)$ and $\theta_2^* = \theta(Q)$ will hold.  Therefore, it only remains to show that $\theta_w^* = \theta(P_{X,T} \otimes Q_{Y \mid X, T})$ under either of the stated assumptions.  In both cases, we will use the following expressions for $\theta(P_{X,T} \otimes Q_{Y \mid X, T})$ and $\theta_w^*$, which are easily derived from standard OLS formulas:
\begin{align*}
&\theta(P_{X,T} \otimes Q_{Y \mid X, T})  = e_1^{\top} \left( \sum_{\ell = 1}^p \E_P[f_{\ell}(X) f_{\ell}(X)^{\top}] \right)^{-1} \sum_{\ell = 1}^p \E_P[ f_{\ell}(X) \tau_{\ell}(X)]\\
  &  \theta_w^* = e_1^{\top} \underbrace{\left( \sum_{\ell = 1}^p \E_Q[ e^{\gamma_0^{* \top} \phi(X)} (\tilde T f_{\ell}(X), g_{\ell}(X))(\tilde T f_{\ell}(X), g_{\ell}(X))^{\top} ] \right)^{-1}}_{:= \bm{\Omega}^{-1}} \underbrace{\sum_{\ell = 1}^p \E_Q[ e^{\gamma_0^{* \top} \phi(X)} Y_{\ell} (\tilde T f_{\ell}(X), g_{\ell}(X))]}_{:= \bm{R}}.
    \end{align*}

First, suppose that $\sum_{\ell = 1}^p f_{\ell}(x) \tau_{\ell}(x) = \bm{B}_0 \phi(x)$ for some matrix $\bm{B}_0$.  We compute $e_1^{\top} \bm{\Omega}^{-1}$.  By the independence of $\tilde T$ and $X$ and the fact that $\tilde T$ is mean-zero, we may conclude that $\bm{\Omega}$ is block-diagonal with upper-left block $\Var_Q(T) \E_Q[ e^{\gamma_0^{* \top} \phi(X)} \sum_{\ell = 1}^p f_{\ell}(X) f_{\ell}(X)^{\top}]$.  Since $\gamma^*$ balances $\phi$ and $\phi$ contains each coordinate of the matrix $\sum_{\ell = 1}^p f_{\ell} f_{\ell}^{\top}$ in its components, this is equal to $\E_P[ \sum_{\ell = 1}^p f_{\ell}(X) f_{\ell}(X)^{\top}]$.  Therefore, we conclude:
    \begin{align}
    e_1^{\top} \bm{\Omega}^{-1} = e_1^{\top} \left( \sum_{\ell = 1}^p \E_P[f_{\ell}(X) f_{\ell}(X)^{\top}] \right)^{-1}. \label{eq:e1Omegainv}
    \end{align}
    Next, we compute the vector $\bm{R}$:
    \begin{align*}
    \bm{R} &= 
    \sum_{\ell = 1}^p \left[ 
    \begin{array}{c}
    \E_Q[ e^{\gamma_0^{* \top} \phi(X)} \E[ Y_{\ell} \tilde T \mid X] f_{\ell}(X)]\\
    \E_Q[ e^{\gamma_0^{* \top} \phi(X)} Y_{\ell} g_{\ell}(X)]
    \end{array}
    \right] &\text{(Definition of $\bm{R}$)}\\
    &= \left[ 
    \begin{array}{c}
    \Var_Q(T) \E_Q [ e^{\gamma_0^{* \top} \phi(X)} \sum_{\ell = 1}^p f_{\ell}(X) \tau_{\ell}(X)]\\
    \sum_{\ell = 1}^p \E_Q[ e^{\gamma_0^{* \top} \phi(X)} Y_{\ell} g_{\ell}(X)]
    \end{array} 
    \right] &\text{(Definition of $\tau_{\ell}$)}\\
    &= \left[ 
    \begin{array}{c}
    \Var_Q(T) \bm{B}_0 \E_Q[ e^{\gamma_0^{* \top} \phi(X)} \phi(X)]\\
    \sum_{\ell = 1}^p \E_Q[ e^{\gamma_0^{* \top} \phi(X)} Y_{\ell} g_{\ell}(X)]
    \end{array}
    \right] &\text{\ref{item:linear_cate}}\\
    &= \left[ 
    \begin{array}{c}
    \Var_Q(T) \bm{B}_0 \E_P[ \phi(X)]\\
    \sum_{\ell = 1}^p \E_Q[ e^{\gamma_0^{* \top} \phi(X)} Y_{\ell} g_{\ell}(X)]
    \end{array}
    \right] &\text{($\gamma^*$ balances $\phi$)}\\
    &= \left[ 
    \begin{array}{c}
    \Var_Q(T) \E_P[ \sum_{\ell = 1}^p f_{\ell}(X) \tau_{\ell}(X)]\\
    \sum_{\ell = 1}^p \E_Q[ e^{\gamma_0^{* \top} \phi(X)} Y_{\ell} g_{\ell}(X)]
    \end{array}
    \right] &\text{\ref{item:linear_cate}}
    \end{align*}
    Putting these two calculations together gives:
    \begin{align*}
    \theta_w^* &= e_1^{\top} \bm{\Omega}^{-1} \bm{R} = e_1^{\top} \left( \Var_Q(T) \sum_{\ell = 1}^p \E_P[ f_{\ell}(X) f_{\ell}(X)^{\top}] \right)^{-1}  \Var_Q(T) \E_P \left[ \sum_{\ell = 1}^p f_{\ell}(X) \tau_{\ell}(X) \right] = \theta(P_{X, T} \otimes Q_{Y \mid X, T}).
    \end{align*}
    Thus we have shown $\Delta^* = \textsf{Covariate shift}$ when \ref{item:linear_cate} holds.

Next, suppose that \ref{item:loglinear_densityratio} holds but \ref{item:linear_cate} may fail.  The proof of \Cref{eq:e1Omegainv} did not use \ref{item:linear_cate}, and hence continues to hold in this case.  Thus, it suffices to compute $\bm{R}$.  It is clear that $(\gamma_0, 0)$ solves the third population moment equation.  Moreover, under a no-collinearity assumption on $\phi$'s components, $(\gamma_0, 0)$ will be the unique solution to this equation.  Hence, $\gamma_0^* = \gamma_0$, allowing us to write:
    \begin{align*}
    \bm{R} &= \sum_{\ell = 1}^p \left[ 
    \begin{array}{c}
    \E_Q[ e^{\gamma_0^{* \top} \phi(X)} \E[ Y_{\ell} \tilde T \mid X] f_{\ell}(X)]\\
    \E_Q[ e^{\gamma_0^{* \top} \phi(X)} Y_{\ell} g_{\ell}(X)]
    \end{array}
    \right] &\text{(Definition of $\bm{R}$)}\\
    &= \sum_{\ell = 1}^p \left[ 
    \begin{array}{c}
    \Var_Q(T) \E_Q[ e^{\gamma_0^{* \top} \phi(X)} \tau_{\ell}(X) f_{\ell}(X)]\\
    \E_Q[ e^{\gamma_0^{* \top} \phi(X)} Y_{\ell} g_{\ell}(X)]\\
    \end{array}
    \right] &\text{(Definition of $\tau_{\ell}$)}\\
    &= \sum_{\ell = 1}^p 
    \left[ 
    \begin{array}{c}
    \Var_Q(T) \E_Q[ \frac{p_X(X)}{q_X(X)} \tau_{\ell}(X) f_{\ell}(X) ]\\
    \E_Q[ e^{\gamma_0^{* \top} \phi(X)} Y_{\ell} g_{\ell}(X) ]
    \end{array}
    \right] &\text{(\ref{item:loglinear_densityratio} and $\gamma_0^* = \gamma_0$)}\\
    &= \left[ 
    \begin{array}{c}
    \Var_Q(T) \E_P[ \sum_{\ell = 1}^p f_{\ell}(X) \tau_{\ell}(X)]\\
    \sum_{\ell = 1}^p \E_Q[ e^{\gamma_0^{* \top} \phi(X)} Y_{\ell}(X) g_{\ell}(X)]
    \end{array}
    \right]
    \end{align*}
    This is the same expression we derived in case \ref{item:linear_cate}, so $\theta_w^* - \theta(P_{X, T} \otimes Q_{Y \mid X, T})$ holds under \ref{item:loglinear_densityratio} as well.
\end{proof}

\subsection{Proof sketch for \Cref{prop:mediation_shift}}
\begin{proof}[Proof sketch for \Cref{prop:mediation_shift}]
By convex duality arguments in \citet{zhao2017entropy}, the weights $\hat{\omega}$ will take the form:
\begin{align*}
\hat{\omega}_i &= \frac{1}{n_2} \exp \left\{ \hat{\gamma}_0^{\top} \phi(X_i) + T_i \hat{\gamma}_1^{\top} \phi(X_i) + \hat{\eta}_0^{\top} \psi(M_i) + T_i \hat{\eta}_1^{\top} \psi(M_i) \right\}.
\end{align*}
Therefore, the same Z-estimator theory referenced in the proof sketch for \Cref{proposition:covariate_shift_properties} implies that the vector $(\theta(\D_1), \beta(\D_1), \theta(\D_2), \beta(\D_2), \hat{\gamma}, \hat{\eta}, \theta_{\omega}(\D_2), \beta_{\omega}(\D_2))$ will have an approximately normal sampling distribution centered at $(\theta_1^*, \beta_1^*, \theta_2^*, \beta_2^*, \gamma^*, \eta^*, \theta_{\omega}^*, \beta_{\omega}^*)$, the solution of the following population estimating equation:
\begin{align*}
\E_{P \times Q} \left[ 
\begin{array}{c}
\sum_{\ell = 1}^p [ Y_{i \ell} - (\theta_1^*, \beta_1^*)^{\top} (\tilde T_i f_{\ell}(X_i), g_{\ell}(X_i))] (\tilde T_i f_{\ell}(X_i), g_{\ell}(X_i))\\
\sum_{\ell = 1}^p [Y_{j \ell} - (\theta_2^*, \beta_2^*)^{\top} (\tilde T_j f_{\ell}(X_j), g_{\ell}(X_j))](\tilde T_j f_{\ell}(X_j), g_{\ell}(X_j))\\
e^{(\gamma^*, \eta^*)^{\top} (\phi(X_j), T_j \phi(X_j), \psi(M_j), T_j \psi(M_j))} \phi(X_j) - \phi(X_i)\\
e^{(\gamma^*, \eta^*)^{\top} (\phi(X_j), T_j \phi(X_j), \psi(M_j), T_j \psi(M_j))} T_j \phi(X_j) - T_i \phi(X_i)\\
e^{(\gamma^*, \eta^*)^{\top} (\phi(X_j), T_j \phi(X_j), \psi(M_j), T_j \psi(M_j))} \psi(M_j) - \psi(M_i)\\
e^{(\gamma^*, \eta^*)^{\top} (\phi(X_j), T_j \phi(X_j), \psi(M_j), T_j \psi(M_j))} T_j \psi(M_j) - T_i \psi(M_i)\\
\sum_{\ell = 1}^p e^{(\gamma^*, \eta^*)^{\top} (\phi(X_j), T_j \phi(X_j), \psi(M_j), T_j \psi(M_j))} [ Y_{j \ell} - (\theta_{\omega}^*, \beta_{\omega}^*)^{\top} (\tilde T_j f_{\ell}(X_j), g_{\ell}(X_j))] (\tilde T_j f_{\ell}(X_j), g_{\ell}(X_j))
\end{array}
\right] = 0.
\end{align*}

Under~\ref{item:linear_cate} or~\ref{item:loglinear_densityratio}, from Proposition~\ref{proposition:covariate_shift_properties} we know that $\theta_w^*$ is consistent and asymptotically normal. With similar reasoning, to prove Proposition~\ref{prop:mediation_shift}, it suffices to show that $\theta_{\omega}^* = \theta(P_{X, T, M} \otimes Q_{Y \mid X, T, M})$ under either of the assumptions stated therein. We begin with some properties of the two quantities. 
Again, as $\tilde{T}$ is centered and independent of $X$, it is straightforward to show that 
\$
\theta(P_{X, T, M} \otimes Q_{Y \mid X, T, M}) 
&= e_1^\top \left( \sum_{\ell = 1}^p \E_P[f_{\ell}(X) f_{\ell}(X)^{\top}] \right)^{-1} \sum_{\ell = 1}^p \E_{P_{X, T, M} \otimes Q_{Y \mid X, T, M}}[ \tilde{T} f_{\ell}(X) Y_\ell] \\ 
&= e_1^\top \left( \sum_{\ell = 1}^p \E_P[f_{\ell}(X) f_{\ell}(X)^{\top}] \right)^{-1} \sum_{\ell = 1}^p \E_{P }\big[ \tilde{T} f_{\ell}(X) \mu_\ell(X,M,\tilde{T} ) \big],
\$
where we recall that $\mu_\ell(x,m,t ) = \EE_{Q}[Y_\ell\given X=x,M=m,\tilde{T}=t]$ is the conditional expectation function under $Q$ (with a non-essential replacement of $T$ by $\tilde{T}$), and $\E_{P_{X, T, M} \otimes Q_{Y \mid X, T, M}}$ denotes the expectation under the reweighted distribution $P_{X, T, M} \otimes Q_{Y \mid X, T, M}$. On the other hand, by definition, $\theta_\omega^* =e_1^\top\bm{\Omega}^{-1}\bm{R}$, where 
\$
\bm{\Omega} &= \sum_{\ell = 1}^p \E_Q\big[ e^{(\gamma^*, \eta^*)^{\top} (\phi(X ), T  \phi(X ), \psi(M ), T  \psi(M ))}  (\tilde T f_{\ell}(X), g_{\ell}(X))(\tilde T f_{\ell}(X), g_{\ell}(X))^{\top} \big],  \\
\bm{R} &= \sum_{\ell = 1}^p \E_Q \big[ e^{(\gamma^*, \eta^*)^{\top} (\phi(X ), T  \phi(X ), \psi(M ), T  \psi(M ))}  Y_{\ell} (\tilde T f_{\ell}(X), g_{\ell}(X))\big] .
\$
Since $(\hat\gamma,\hat\eta)$ balances all coordinates in the matrix $\sum_{\ell=1}^p f_\ell f_\ell^\top$, we know that 
\#
e_1^\top \bm{\Omega}^{-1} =  e_1^\top \left( \sum_{\ell = 1}^p \E_P[f_{\ell}(X) f_{\ell}(X)^{\top}] \right)^{-1}.
\#
This is regardless of the properties of the outcomes and the density ratios. Furthermore, by the design of the balancing features, $\hat{w}$ essentially balances $\phi(X)$ and $\psi(M)$ separately in the treatment group $(T=1)$ and control group $(T=0)$.

We now proceed to show the consistency of $\theta_\omega^*$ under either of the assumptions in Proposition~\ref{prop:mediation_shift}.

First, suppose~\ref{item:linear_outcome_model_mediator} holds, i.e., $\sum_{\ell = 1}^p f_{\ell}(x) \mu_{\ell}(x, m, t) = \bm{U}_t \phi(x) + \bm{V}_t \psi(m)$. 
Then, 
\$
\bm{R} &= 
\begin{bmatrix}
\sum_{\ell = 1}^p \E_Q \big[ e^{(\gamma^*, \eta^*)^{\top} (\phi(X ), T  \phi(X ), \psi(M ), T  \psi(M ))}  Y_{\ell}  \tilde T f_{\ell}(X) \big] \\[0.5ex]
\sum_{\ell = 1}^p \E_Q \big[ e^{(\gamma^*, \eta^*)^{\top} (\phi(X ), T  \phi(X ), \psi(M ), T  \psi(M ))}  Y_\ell g_{\ell}(X) \big]
\end{bmatrix} &\text{(definition of $\bm{R}$)} \\ 
&= \begin{bmatrix}
\sum_{\ell = 1}^p \E_Q \big[ e^{(\gamma^*, \eta^*)^{\top} (\phi(X ), T  \phi(X ), \psi(M ), T  \psi(M ))}  \tilde T f_\ell(X)\mu_\ell(X,M,\tilde{T}) \big] \\[0.5ex]
\sum_{\ell = 1}^p \E_Q \big[ e^{(\gamma^*, \eta^*)^{\top} (\phi(X ), T  \phi(X ), \psi(M ), T  \psi(M ))} Y_\ell  g_{\ell}(X) \big]
\end{bmatrix} &\text{(tower property)} \\ 
&= \begin{bmatrix}
\sum_{\ell = 1}^p \E_Q \big[ e^{(\gamma^*, \eta^*)^{\top} (\phi(X ), T  \phi(X ), \psi(M ), T  \psi(M ))}  \tilde T  f_{\ell}(X) (\bm{U}_T \phi(X) + \bm{V}_T \psi(M)) \big] \\[0.5ex]
\sum_{\ell = 1}^p \E_Q \big[ e^{(\gamma^*, \eta^*)^{\top} (\phi(X ), T  \phi(X ), \psi(M ), T  \psi(M ))} Y_\ell  g_{\ell}(X) \big]
\end{bmatrix} &\text{\ref{item:linear_outcome_model_mediator}} \\ 
&= \begin{bmatrix}
\sum_{\ell = 1}^p \E_P \big[ \tilde T  f_{\ell}(X) (\bm{U}_T \phi(X) + \bm{V}_T \psi(M)) \big] \\[0.5ex]
\sum_{\ell = 1}^p \E_Q \big[ e^{(\gamma^*, \eta^*)^{\top} (\phi(X ), T  \phi(X ), \psi(M ), T  \psi(M ))} Y_\ell  g_{\ell}(X) \big]
\end{bmatrix} &\parbox{8em}{($\gamma^*$ and $\eta^*$ balances $\psi(M)$ and $\phi(X)$ in two treatment groups)} \\
&= \begin{bmatrix}
\sum_{\ell = 1}^p \E_P \big[ \tilde T  f_{\ell}(X) \mu_\ell(X,M,\tilde{T}) \big] \\[0.5ex]
\sum_{\ell = 1}^p \E_Q \big[ e^{(\gamma^*, \eta^*)^{\top} (\phi(X ), T  \phi(X ), \psi(M ), T  \psi(M ))} Y_\ell  g_{\ell}(X) \big]
\end{bmatrix} .&\text{\ref{item:linear_outcome_model_mediator}}  \\
\$
Putting the two pieces together, we obtain 
\$
e_1^\top\bm{\Omega}^{-1}\bm{R} =  e_1^\top \left( \sum_{\ell = 1}^p \E_P[f_{\ell}(X) f_{\ell}(X)^{\top}] \right)^{-1} \sum_{\ell = 1}^p \E_P \big[ \tilde T  f_{\ell}(X) \mu_\ell(X,M,\tilde{T}) \big] = \theta(P_{X, T, M} \otimes Q_{Y \mid X, T, M}).
\$

Second, suppose~\ref{item:loglinear_densityratio_mediator} holds but~\ref{item:linear_outcome_model_mediator} is likely to fail. Recall that $\hat{\omega}$ balances $\psi(M)$ and $\phi(X)$ simultaneously in the treated and control groups. By the same argument as for the covariate shift component, $\gamma^*$ and $\eta^*$ are unique solutions to the moment equations, and they correspond to the correct density ratios in the two groups. Inside each treatment arm, the mediators $M$ can be viewed as plain covariates. Thus, similar to the case~\ref{item:loglinear_densityratio}, we can show that $\sum_{\ell = 1}^p \E_Q \big[ e^{(\gamma^*, \eta^*)^{\top} (\phi(X ), T  \phi(X ), \psi(M ), T  \psi(M ))}  Y_{\ell}  \tilde T f_{\ell}(X) \big] = \sum_{\ell = 1}^p \E_P \big[ \tilde T  f_{\ell}(X) \mu_\ell(X,M,\tilde{T}) \big] $, which further leads to $e_1^\top\bm{\Omega}^{-1}\bm{R}  = \theta(P_{X, T, M} \otimes Q_{Y \mid X, T, M}).$
We thus conclude the proof of Proposition~\ref{prop:mediation_shift}.
\end{proof}

\subsection{Proof sketch for \Cref{lemma:truncated_normal}}
\begin{proof}[Proof sketch for \Cref{lemma:truncated_normal}]
By Bayes rule:
\begin{align*}
&\P( \hat{\Delta} \in A \mid \text{Original study published/replicated with }\hat{p} < \alpha_0)\\
&\quad = \frac{\P( \text{Original study published/replicated}, \hat{p} < \alpha_0, \hat{\Delta} \in A)}{\P( \text{Original study published/replicated}, \hat{p} < \alpha_0)}\\
&\quad = \frac{\P( \hat{p} < \alpha_0) \P( \text{Original study published/replicated}, \hat{\Delta} \in A \mid \hat{p} < \alpha_0)}{\P( \text{Original study published/replicated}, \hat{p} < \alpha_0)}\\
&\quad = \frac{\P( \hat{p} < \alpha_0) \P( \text{Original study published/replicated} \mid \hat{p} < \alpha_0) \P( \hat{\Delta} \in A \mid \hat{p} < \alpha_0)}{\P( \text{Original study published/replicated}, \hat{p} < \alpha_0)} &\text{(\Cref{assumption:selection})}\\
&\quad = \P( \hat{\Delta} \in A \mid \hat{p} < \alpha_0)\\
&\quad = \P( \hat{\Delta} \in A \mid | \hat{\Delta}_1| > z_{1 - \alpha_0/2})\\
&\quad \approx \P_{Z \sim N(\Delta ,\bm{\Sigma})}( Z \in A \mid |Z_1| > z_{1 - \alpha_0/2}).
\end{align*}
We remark that in the last line, we used the approximation of the \emph{marginal} distribution to conclude approximation of the \emph{conditional} distribution.  This is valid only for conditioning events with non-negligible probability.  However, the event $\{ |\hat{\Delta}_1| > z_{1 - \alpha_0/2} \}$ is substantial enough for this to be valid.
\end{proof}



\end{document}